\RequirePackage{fix-cm}
\documentclass[natbib,smallextended]{svjour3}       
\bibpunct{[}{]}{;}{n}{}{,} 

\smartqed  
\usepackage{graphicx}
\usepackage{color}
 \journalname{my journal}
\newcommand {\nc} {\newcommand}
\nc {\beq} {\begin{eqnarray}}
\nc {\eeq} {\nonumber \end{eqnarray}}
\nc {\eeqn} [1] {\label{#1} \end{eqnarray}}
\nc {\eol} {\nonumber \\}
\nc {\rref} [1] {(\ref{#1})}
\nc{\ETAL} {\mbox{\sl et al.}}
\nc {\vR} {\mbox{$\ve{R}$}}
\nc {\ve} [1] {\mbox{\boldmath $#1$}}
\nc {\la} {\mbox{$\langle$}}
\nc {\ra} {\mbox{$\rangle$}}
\nc {\cL} {\mbox{${\cal L}$}}
\nc {\dem} {\mbox{$\frac{1}{2}$}}
\nc {\arrow} [2] {\mbox{$\mathop{\rightarrow}\limits_{#1 \rightarrow #2}$}}
\nc {\wiggle} [2] {\mbox{$\mathop{\sim}\limits_{#1 \rightarrow #2}$}}
\nc {\red}[1] {\textcolor{red}{#1}}
\begin{document}

\title{Hyperspherical cluster model for bosons: application to  
sub-threshold halo states in helium drops
}

\titlerunning{Hyperspherical cluster model for bosons ...
}        

\author{N.K. Timofeyuk
}


\institute{ N.K. Timofeyuk \at
              Physics Department, University of Surrey, Guildford,
Surrey GU2 7XH, England, UK \\
\email{ n.timofeyuk@surrey.ac.uk}
}

\date{Received: date / Accepted: date}

\maketitle

\begin{abstract}

To describe long-range behaviour of one particle removed from a few- or a many-body system, a hyperspherical cluster model has been  developed. It has been applied to  the ground and   first excited states of  helium drops with five, six, eight and ten atoms interacting via a two-body soft gaussian potential. Convergence of the hyperspherical cluster harmonics expansion is studied for binding energies, root-mean-squared radii and  overlaps of the wave functions of two helium drops differing by one atom. It was shown that with increasing model space  the functional form of such overlaps at large distances converges to the correct asymptotic behaviour. The asymptotic normalization coefficients that quantify the overlaps' amplitudes in this region are calculated.
It was also shown that in the first excited state one helium atom stays far apart from the rest forming a two-body molecule, or a halo. The probability of finding the halo atom in the classically-forbidden region of space depends on the definition of the latter and on the number of atoms in the drop.  The total norm of the overlap integrals,  the spectroscopic factor,  represents the number of partitions of a many-body state  into a chosen state of the system with one particle removed. The spectroscopic factors have been calculated and their sum rules are discussed giving a further insight into the structure of helium drops.

\end{abstract}

\section{Introduction}
\label{intro}

Most methods of solving the Schr\"odinger equation for few-  and many-body systems employ  a wave function expansion over some sets of  basis states. Choosing   hyperspherical harmonics (HH) as a   basis 
reduces the Schr\"odinger equation  to a system of coupled equations of only one dynamic variable for any number $A$ of particles \cite{SSh}, making it a powerful tool for the studies of  quantum systems.
The HH expansion is routinely used to describe various processes in few-body nuclear \cite{Mar20} and atomic \cite{Avery} systems. It has also been applied to heavier systems as well, see, for example, \cite{SSh, Bar99, Ada99, Tim08a, Kie17}. The drawback of any basis-expansion method is that as the number of basis states grows with the number of particles in the systems  the problem of solving the Schr\"odinger equation becomes unmanageable. However, one can expect that not all basis states are important for describing selected properties of an $A$-body systems. Therefore, an efficient way to reduce the number of basis states is needed. Selecting the most important of them depends on the observables of interest to be calculated. 

The primary interest of all few- and many-body calculations are binding energies. For short-range pairwise interactions the  energies of tightly-bound systems depend on how good their wave functions are described in the  internal region. However, most interesting physical phenomena occur when different  systems react with each other and go though various rearrangements. Such processes are determined by how good the wave function of the total system is described at large separations between its different subsystems. 
To deal with long-range behaviour,  a hyperspherical cluster model (HCM) has been proposed  for fermionic systems in Ref. \cite{Tim07}. The HCM    selects specific HH states responsible for the behaviour of one nucleon staying in a classically forbidden region. The model has been tested for a toy $^5$He case where this nucleus was artificially made bound. The projection of the $^5$He wave function   onto the $^4$He wave function was constructed proving that it converges towards a correct asymptotic form with increasing number of the HCM basis states. However, in a particular method  of calculating the hyperradial potentials chosen in \cite{Tim07} the increasing model space led to a need for summation of many terms differing by more than 16 orders of magnitude, inducing large uncertainties in the calculations and preventing further use of HCM. 

 In this paper, the HCM has been revisited both formally and computationally and reformulated to avoid the problems of its original version. To focus on the coordinate part of the wave function only, it has been worked out  for  bosonic systems with zero spin, with numerical applications to helium drops. There is a considerable interest to these systems because of the universality of some of their properties, 
 arising due to the extremely weak binding in the He-He dimer. In this dimer the atoms stay in a classically forbidden region, far apart from each other. The spectrum of the helium trimer is a good example of the manifestation of Efimov physics \cite{Nai17} that deals with systems composed of weakly-interacting constituents.
 The helium trimer has one deep and one shallow bound state with the ratio of their energies being close to the universal value predicted by Efimov \cite{Efimov}. The shallow Efimov first excited state has been experimentally observed in \cite{Kun15} and  it   consists of a dimer to which the third helium atom is weakly attached and located at even larger distance than two atoms in the dimer \cite{Kun22}. In  \cite{Gat14,Kie14a,Kie14b} it was shown   that the existence of a shallow excited state below the one-atom removal threshold persists in helium drops with more than three atoms. 
One can expect that in such states one helium atom would also stay far away from the rest, spreading out into a classically forbidden region and forming a two-body molecule. In nuclear physics, this phenomenon is called  "halo" \cite{Rii94} and it has been extensively studied using few-body models \cite{Bha21}. The HCM, designed to treat single-particle long-range behaviour, should be  ideal for studying halo states.

To find out if halo persists in the first excited states of helium drops with increasing number of atoms, the overlap integrals between $A$ and $A-1$-body wave functions were constructed. Such overlaps are common in nuclear physics  where they serve as an input for  nucleon-removal-reaction calculations \cite{Satchler}. In atomic physics they are called Dyson orbitals \cite{Ort04} and they are used in photoelectron spectroscopy and $(e,2e)$ ionization experiments.
The overlap integrals, calculated for helium drops, should   quantify  spatial behaviour of a single atom in the whole system, giving an idea about the size of the two-body molecule, while their norms 
give the spectroscopic strength of the molecular states.

This paper starts with definition of  the hyperspherical harmonics expansion in section \ref{sec:2} followed by the hyperspherical cluster harmonics basis construction in 
section \ref{sec:3}. The HCM applications below use a soft gaussian potential for which convergence of the ground-state binding energies dramatically improves with increasing number of atoms in the drop \cite{Tim12}. Therefore,  the lowest-order approximation is used to describe the $A-1$-core of the $A$-body system, which is described in section \ref{sec:4}.  Section \ref{sec:5} presents numerical results for binding energies and root-mean-squared radii of helium drops with $A=5,6,8,10$, while section \ref{sec:6} discusses their overlap integrals calculations.
Section \ref{sec:7} contains a summary of the work and concluding remarks while the Appendix provides information about important but lengthy mathematical details relevant to the HCM development.

\section{Hyperspherical harmonics expansion and Schr\"odinger equation}
\label{sec:2}

Hyperspherical coordinates could be chosen in many possible ways. Here we will follow a canonical choice based on the set of $A-1$ normalised Jacobi coordinates $\ve{\xi}_1,\ve{\xi}_2,\dots,\ve{\xi}_{A-1}$ defined as,
\beq
\ve{\xi}_{A-i}=\sqrt{\frac{i}{i+1}} \left(\frac{1}{i} \sum_{j=1}^i \ve{r}_j - \ve{r}_{i+1} \right),
\eeqn{Jacobi}
where $\ve{r}_ i$ is the  individual coordinate of the $i$-th body. This definition uses $\ve{\xi}_{A-i}$ in the l.h.s. of (\ref{Jacobi}) instead of the standard choice  of $\ve{\xi}_i$  to simplify notations in the sections below. 
The Jacobi coordinates form the $(A-1)$-dimensional vector $\ve{\rho}$ with the length   given by the hyperradius $\rho$,
\beq
\rho^2 = \sum_{i=1}^{A-1} \ve{\xi}_i^2 =
\sum_{i=1}^A \ve{r}_{ i}^2 -  \ve{R}^2  =
\frac{1}{A}\sum_{i < j}^A(\ve{r}_{i}-\ve{r}_{ j})^2  ,
\eeqn{f1}
where $\ve{R} = (\sum_{i=1}^A \ve{r}_{i})/\sqrt{A}$ is the normalised  coordinate of the centre of mass. 
The other hyperspherical coordinates are the hyperangles $\hat{\rho} \equiv \{\theta_{3A-4}, \, \theta_{3A-3},$ $\dots,\theta_{2}, \theta_1\}$ and in HCM the last two hyperangles are chosen as follows
\beq
\xi_{1} &=& \rho \cos \theta_{1},
\eol
\xi_{2} &=& \rho \sin\theta_1 \cos \theta_2.
\eeqn{}

The kinetic-energy operator $\hat{T}$ in hyperspherical coordinates consists of the hyperradial and hyperangular parts,
\beq
\hat{T} = -\frac{\hbar^2}{2m} \left(\frac{1}{\rho^{n-1}} \frac{\partial}{\partial \rho}
\left(\rho^{n-1} \frac{ \partial}{\partial\rho}\right) - \frac{ 1}{ \rho^2}
\Delta_{\hat{\rho}}
\right),
\eeqn{f7a}
where  $m$ is the mass of the boson and $n = 3A -3$ is the dimension of space formed by Jacobi coordinates. 
The eigenfunctions $Y_{K\gamma}(\hat{\rho})$ of the hyperangular part $\Delta_{\hat{\rho}}$, 
\beq
\Delta_{\hat{\rho}}Y_{K\gamma}(\hat{\rho}) = K(K+n-2) Y_{K\gamma}(\hat{\rho}),
\eeqn{f4}
where $K$ is the hypermomentum,  
form a complete orthonormal set, calles hyperspherical harmonics (HH). 
They can be highly degenerate and, therefore, different harmonics belonging to the same $K$ are labelled by $\gamma$.
For bosonic systems, HHs are constructed symmetrical with respect to particle permutations.

After expanding the wave function of the $A$-body system, $\Psi(\ve{\rho})$, 
over complete HH sets,
\beq
\Psi(\ve{\rho})=
\rho^{-\frac{n-1}{2}} \sum_{K\gamma} \chi_{K\gamma}(\rho) Y_{K\gamma}(\hat{\rho}),
\eeqn{expansion}
one gets the infinite set of coupled differential equations for the radial parts $ \chi_{K\gamma}(\rho) $:
\beq
 \left(-\frac{d^2}{d\rho^2}+\frac{{\cal L}_{K}({\cal L}_{K}+1)}{\rho^2}
-\frac{2m}{\hbar^2} \left(E - V_{K\gamma,K\gamma}(\rho)\right) \right)
\chi_{K\gamma}(\rho)
\eol
= -\frac{2m}{\hbar^2}
\sum_{K'\gamma'\ne K\gamma} V_{K\gamma_,K'\gamma'}(\rho) \chi_{K'\gamma'}(\rho).
\eeqn{f5}
Here   
\beq
{\cal L}_{K} = K + \frac{n-3}{2}
\eeqn{LK}
is the generalized angular momentum, $m$ is the mass of the particle
and the hyperradial potentials $V_{K\gamma,K'\gamma'}(\rho)$ are 
the matrix elements of the interaction potential $\hat{V}$ between HHs,
\beq
V_{K\gamma,K'\gamma'}(\rho) =
\la Y_{K\gamma}(\hat{\rho}) | \hat{V}|
Y_{K'\gamma'}(\hat{\rho}) \ra.
\eeqn{f6}

\section{Construction of the hyperspherical cluster basis states.}
\label{sec:3}

The hyperspherical cluster harmonics (HCH) states will be constructed below following the suggestion in \cite{Tim07} to reorganize the HH states to the form   resembling the cluster wave functions used by microscopic cluster models \cite{Des12}.
This is achieved by symmetrizing the product of a (known)
completely symmetric HH
$Y_{K_c\gamma_c} (\hat{\ve{\rho}}_c)$ 
for the core $A-1$, which is a function of the hyperangles 
$\hat{\ve{\rho}}_c$,  
and  a relative hyperangular function
$\varphi_{K_c \nu lm} (\theta_1,\hat{\xi}_{1})$
\beq
{\cal Y}_{K \gamma}(\hat{\ve{\rho}})  = 
{\cal N}_{K \gamma}^{-1}
\,\,{\cal S}  
\left[Y_{K_c\gamma_c} (\hat{\ve{\rho}}_c)\,\,
\times \varphi_{K_c\nu l} 
(\theta_1,\hat{\xi}_{1})\right]_{LM}
\eeqn{defY}
Here  $K_c$ is the hypermomentum,
$\gamma_c = \{\beta_c L_c M_c\}$, $L_c$
are the total orbital momentum
of the core $A-1$ while the $\beta_c$ labels  different degenerate harmonics belonging to the same $K_c$. The set of quantum numbers $K\gamma$ contains $\gamma = \{K_c\gamma_c \nu l\,LM\}$. The core HH and the relative hyperangular functions are coupled into the total orbital moment $\ve{L} = \ve{L}_c + \ve{l}$.
Also, ${\cal N}_{K \gamma}$ is a normalization factor and
\beq
{\cal S} = \frac{1}{A^{1/2}} \left(1 + \sum_{i=1}^{A-1} P_{iA}\right)
\eeqn{S}
 is the symmetrisation operator
which permutes the $A$-th boson with bosons of the $A-1$  core. The relative HH, associated with the last nucleon, is given by the expression \cite{SSh}
\beq
\varphi_{K_c \nu lm}^{(n)} (\theta,  \hat{{\xi}})=  N^{(n)}_{ K_c\nu l} \,
\cos^l\theta  \sin^{K_c} \theta \,
P_{\nu}^{K_c+\frac{n-5}{2},l+\frac{1}{2}}(\cos 2\theta)
Y_{lm}(\hat{ { \xi}}),
\eeqn{varpi}
where 
the normalization is given by
\beq
N_{nK_cl}^{(n)} =  \left(
\frac{2\nu!\,(2\nu+K_c+l+(n-2)/2)\Gamma(\nu+K_c+l+(n-2)/2)}
{ \Gamma(\nu+l+3/2) \,\Gamma(\nu+K_c+(n-3)/2)}\right)^{1/2}. \,\,\,
\eeqn{nkkl}
 
Since  $Y_{K_c\gamma_c}(\hat{\rho}_c)$
and  $\varphi_{K_c\nu lm}(\theta,\hat{ {\xi}}_{1})$ are
 the eigenfunction of the operator of the
kinetic energy in subspaces associated with variables 
$\hat{\ve{\rho}}_c$ and $\{\theta_1,\hat{{\xi}}_{1}\}$ and because 
both the operator of the
kinetic energy  and the hyperangular function
(\ref{defY}) are symmetric with respect to any bosonic permutations,
the function ${\cal Y}_{K \gamma}(\hat{\ve{\rho}})$
is also an eigenfunction of  $\Delta_{\hat{\rho}}$, satisfying Eq.   (\ref{f4}),
 and therefore it is the HH with the hypermomentum
$K = K_c+2\nu+l$. The HCHs 
${\cal Y}_{K\gamma}(\hat{\ve{\rho}})$ with different values of $K$ are orthogonal to each other, however, for the same
$K$ and different $\gamma$ they may be not orthogonal
\beq
\la {\cal Y}_{K'\gamma'}(\hat{\ve{\rho}})|{\cal Y}_{K\gamma}(\hat{\ve{\rho}})\ra  
= \delta _{KK'} {\cal I}_{K \gamma \gamma'}
,
\eeqn{yy}
which is common for different channel functions in multichannel cluster models with antisymmetrization. In this case orthogonalization could be carried out.

In Ref. \cite{Tim07} the  HCM hyperradial potentials  $V_{K\gamma,K'\gamma'}(\rho)$  were  constructed   based on the link between the HCHs and  the basis states of the translation-invariant harmonics oscillator cluster model. The formalism of \cite{Tim07} used fractional parentage expansions of the harmonic oscillator wave functions of the  $A-1$ core
 followed by  re-expressing   the products of harmonic oscillator wave functions in  the coordinates that are most convenient  for performing analytical integrations. Here,  harmonic oscillator representation of HCHs is not used. Instead,  fractional parentage expansion of the core harmonic $Y_{K_c \gamma_c}(\hat{\rho}_c)$ is employed and then coordinate transformations involving hyperangles $\theta_1$ and $ \theta_2$ are performed. Although techniques to calculate fractional parentage expansion coefficients are available, see  \cite{SSh} or \cite{Bar97}, they are rarely used in modern few- and many-body applications. Adapting these techniques for the needs of HCM requires additional effort. However, because the aim of the present work is to construct an approach to treat  long-range behavior only, as a starting point, the case of $K_c=0$   is considered here. With this choice, and for soft two-body potentials considered below,   the $A-1$ core binding energy  does not differ much from a converged result, progressively getting closer to it with increasing number of bosons \cite{Tim12}. Since the HH for $K_c=0$   is just a constant, its fractional parentage expansion is trivial. Separating one boson from the rest results in factorising $Y_{K_c=0}(\hat{\rho}_c)$ into the product  $\prod_{i=1}^{A-2}\varphi^{(3i)}_{000}(\theta_i, \hat{\xi}_i)$ of the functions $\varphi_{000}$ given, by eq. (\ref{varpi}), which   are  all constants.


\section{The $K_c=0$  case}
\label{sec:4}


We consider the case of   $K_c = 0$, $L_c =0$ and, in addition, we will keep $l=0$ as well to describe the ground and first excited 0$^+$ states.  In this case
\beq
{\cal Y}_{\nu}(\hat{\ve{\rho}}) =  {\cal N}^{-1}_{\nu} {\cal S} \left[ Y_0(\hat{\rho}_c) \varphi^{(n)}_{0 \nu 0} (\theta_1, \hat{\xi}_1)\right].
\eeqn{HCH0}
With this choice for a fixed value of $K$ there is only one HCH available with $K = 2\nu$ and the set (\ref{HCH0}) is orthonormal, $\la {\cal Y}_{\nu}(\hat{\ve{\rho}}) \mid {\cal Y}_{\nu'}(\hat{\ve{\rho}}) \ra= \delta_{\nu,\nu'}$.

It is well known that for bosonic systems the HHs with $K=2$ could not be constructed. It is easy to verify that for $\nu=1$ 
\beq
{\cal S} \left[ { Y}_0 (\hat{\vec{\rho}}_c)  \,\phi_{010}(A)\right]\sim \left(\frac{n}{2}\frac{A}{A-1}-\frac{3}{2}A\right){Y}_0 (\hat{\vec{\rho}}),  
\eeqn{y1=0}
which gives   ${\cal Y}_1 = 0$. Therefore, in the HCH expansion of the bosonic wave function the $K=2$ term will be missing.

\subsection{Calculating the HCH norm}
Because in the $K_c=0$ case the HCHs are not degenerate, the norm  ${\cal N}_{\nu}$ is calculated from $\la {\cal Y}_{\nu}(\hat{\vec{\rho}}) | {\cal Y}_{\nu}(\hat{\vec{\rho}})\ra =1$, which after separating   $\varphi^{(n-3)}_{000}(\theta_2,\hat{\xi}_{2})$ from $Y_0(\hat{\rho}_c)$ takes form
\beq
& &\la {\cal Y}_{\nu}(\hat{\ve{\rho}}) | {\cal Y}_{\nu}(\hat{\ve{\rho}})\ra 
=
{\cal N}^{-2}_{ \nu }\left( 1 + (A-1)
\right. \eol &\times&  \left.  \la \, \varphi^{(n-3)}_{000} 
(\theta'_2,\hat{{\xi}}'_{2}) \varphi^{(n)}_{0\nu 0} (\theta'_1,\hat{\xi}'_{1}) 
\mid
 \varphi^{(n-3)}_{000 } (\theta_2,\hat{{\xi}}_{2})   \varphi^{(n)}_{0\nu 0 } 
(\theta_1,\hat{\xi}_{1})  \ra \right).
\eeqn{}
The vectors $\vec{\xi}'_2,\vec{\xi}'_1 $ are shown in Fig.1 together with  $\vec{\xi}'_2,\vec{\xi}'_1 $. It is shown in the Appendix that for the coordinate transformation
\beq
\vec{\xi}_1 &=& \gamma_1 \vec{\xi}'_1 +\gamma_2 \vec{\xi}'_2 \eol
\vec{\xi}_2 &=& -\gamma_2 \vec{\xi}'_1 +\gamma_1 \vec{\xi}'_2 
\eeqn{c-trans}
with $\gamma_1^2 + \gamma^2_2=1$ the product $ \varphi^{(n-3)}_{000 } (\theta_2,\hat{{\xi}}_{2})   \varphi^{(n)}_{0\nu 0 } 
(\theta_1,\hat{\xi}_{1}) $ could be expressed via the product of eigenfunctions expressed in  coordinates $(\theta'_1, \hat{\xi}'_1,\theta'_2, \hat{\xi}'_2)$ as
\beq
\varphi^{(n-3)}_{000 } (\theta_2,\hat{{\xi}}_{2})   \varphi^{(n)}_{0\nu 0 } 
(\theta_1,\hat{\xi}_{1}) 
&=& \sum_{\nu_1+\nu_2+l=\nu} T^{(n)}_{\nu,\nu_1 \nu_2 l} (\gamma_1,\gamma_2)
\eol &\times& 
[\varphi^{(n-3)}_{0\nu_2l } (\theta'_2,\hat{{\xi}}'_{2})  \times \varphi^{(n)}_{2\nu_2+l \, \nu l } 
(\theta'_1,\hat{\xi}'_{1})]_0 .\,\,\,\,
\eeqn{TC}
The hyperangular functions in new coordinates may have different angular momentum but they are coupled to the total angular momentum equal to zero. The coefficients $ T^{(n)}_{\nu,\nu_1 \nu_2 l} (\gamma_1,\gamma_2)$ are analogs of well-known Raynal-Revai coefficients \cite{RR} used for calculating the hyperradial potentials in three-body problems.  They satisfy
\beq
\sum_{\nu_1+\nu_2+l=\nu} T^{(n)}_{\nu,\nu_1 \nu_2 l} (\gamma_1,\gamma_2)T^{(n)}_{\nu',\nu_1 \nu_2 l} (\gamma_1,\gamma_2) = \delta_{\nu,\nu'}.
\eeqn{sumT}
For orthonormal HHs
one gets
\beq
{\cal N} _{ \nu } = \left[ 1 + (A-1)T^{n}_{\nu,  \nu 00}(\gamma_1^{{\cal N}},\gamma_2^{{\cal N}})
\right]^{1/2}
\eeqn{}
with
\beq
 \gamma^{{\cal N}}_1 = -\frac{1}{A-1} , 
 \,\,\,\,\,\,\,\,\,\,\,\,
 \gamma^{{\cal N}}_2 =  \sqrt{\frac{A(A-2)}{(A-1)^2}}.
 \eeqn{gam_norm_N}
It is easy to check that for $\nu=1$
\beq
1+(A-1)T^{(n)}_{1,100}\left(\gamma^{{\cal N}}_1,\gamma^{{\cal N}}_2\right) = 0,
\eeqn{N1}
so that the norm of ${\cal Y}_1$ is not defined.
 
 \begin{figure}[t]
\vspace{0.5 cm}
\includegraphics[scale=0.38]{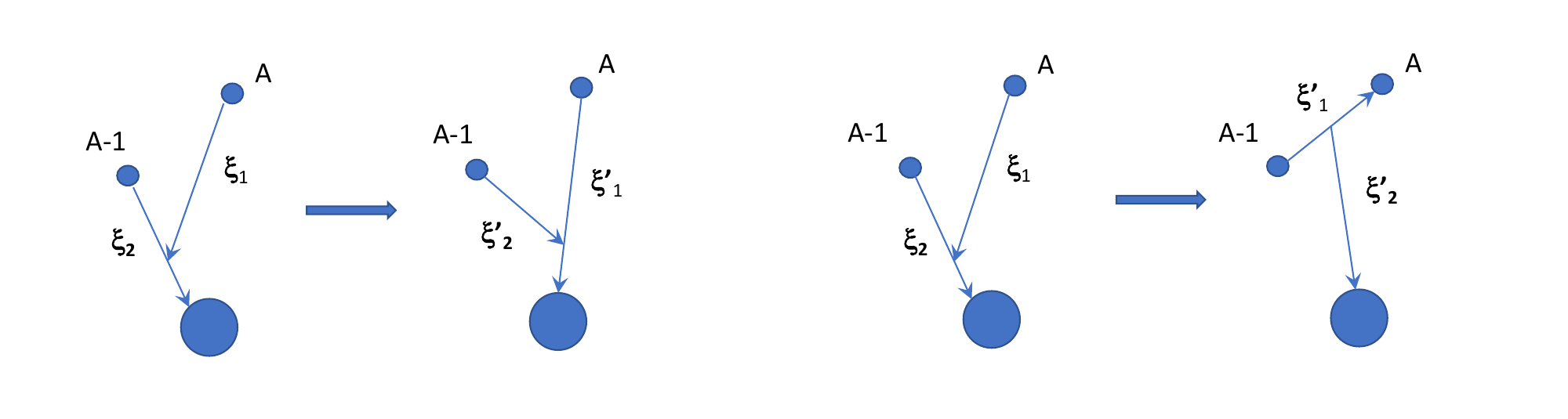}
\caption{Coordinate transformations used for calculating the HCH norm (left) and direct contribution to hyperradial potential from interaction between boson $A$ and $A-1$ (right).}
\label{fig:V4ex}
\end{figure}

\subsection{Calculating the hyperradial potential }

The  matrix element  
$ \la {\cal S} \psi_{\alpha'}|V|{\cal S}\psi_{\alpha} \ra 
$
between arbitraty   wave functions $\psi_{\alpha}=\varphi_{\alpha_1}(1,2,...,A-1)\varphi_{\alpha_2}(A)$, containing symmetric wave function $\varphi_{\alpha_1}(1,2,...,A-1)$ of the core $A-1$, can be split into  direct and exchange terms,
\beq
\la {\cal S} \psi_{\alpha'}|V|{\cal S}\psi_{\alpha} \ra 
= V_{dir}+(A-1)V_{ex}.
\eeqn{arb1}
The direct term 
\beq 
V_{dir} =\la \phi_{\alpha_1'}(1,...,A-1) 
\varphi_{\alpha_2'}(A)|V|\phi_{\alpha_1}(1,...,A-1)\varphi_{\alpha_2}(A)\ra
\eeqn{dir}
can be separated into two parts,
\beq
V_{dir} &=& V_{dir}^{(1)} + V_{dir}^{(2)} 
\eol
&= & 
\la \phi_{\alpha_1'}(1,...,A-1)\varphi_{\alpha'_2}(A)|\sum_{i<j}^{A-1}V_{ij}|\phi_{\alpha_1}(1,...,A-1)\varphi_{\alpha_2}(A)\ra 
\eol
 &+&
\la \phi_{\alpha_1'}(1,...,A-1) \varphi_{\alpha_2'}(A)|\sum_{i=1}^{A-1}V_{iA}|
\phi_{\alpha_1}(1,...,A-1)\varphi_{\alpha_2}(A)\ra, 
\eeqn{Vdir}
the first of which, $V_{dir}^{(1)}$, contains the interaction between
  $A-1$ particles in the core while
  the second term $V_{dir}^{(2)}$   involves interactions between the   particle $A$ and the particle of the core $A-1$. 
The exchange potential, 
\beq
V_{ex} =
\la \phi_{\alpha_1'}(1,...,A) 
\varphi_{\alpha_2'}(A-1)|V|\phi_{\alpha_1}(1,...,A-2,A-1)
\varphi_{\alpha_2}(A)\ra
\eeqn{ex} 
can be split into four terms,
\beq
V_{ex} = V_{ex}^{(1)} + V_{ex}^{(2)} + V_{ex}^{(3)} + V_{ex}^{(4)},  
\eeqn{vex}
according to the following separation of the two-body interaction potential:
\beq
\sum_{i<j}^A V_{ij} = V_{A-1,A} + \sum_{i=1}^{A-2} V_{iA}+ 
\sum_{i=1}^{A-2} V_{iA-1} +\sum_{i<j}^{A-2} V_{ij}.
\eeqn{}
The HCM hyperradial coupling potentials
\beq
V_{\nu \nu'}(\rho) = \la {\cal Y}_{\nu}(\hat{\ve{\rho}})  \mid \sum_{i < j} V_{ij}  \mid {\cal Y}_{\nu'}(\hat{\ve{\rho}}) \ra
\eeqn{VHR}
are then given by six terms,
the formal expressions for which are rather cumbersome, therefore, they are moved to Appendix. To validate numerical calculations it was  checked that replacing the two-body gaussian potential either by a constant or a two-body harmonic oscillator interaction results in $V_{\nu'\nu}(\rho)= \frac{A(A-1)}{2}\delta_{\nu,\nu'}$. It was also verified that for a gaussian interaction $V_{1,\nu}(\rho) = 0$ for all $\nu$. The calculation of one-dimensional integrals over a hyperspherical angle, which are a part of (\ref{VHR}) and are shown in Appendix, a Gauss-Jacobi rule was used with the help of the subroutine from \cite{gaujac}. While only 10 points were needed to get a good accuracy of hyperradial potentials for ground state binding energy calculations, to get a similar accuracy for the hyperradial wave functions at very large distances the number of point was increased to 30.

\section{Numerical applications to the binding energies and radii of helium drops}
\label{sec:5}

\begin{table*}
\caption{ Absolute values of binding energies (in K) for the  ground and first excited 0$^+$ states of a system of $A$ helium atoms
calculated  for several values of $K_{\max}=2\nu_{\max}$ in the full HH basis and in the HCM.
}
      \begin{tabular}{p{0.7 cm}p{0.9 cm}p{1.0 cm}p{0.8 cm}p{1.1cm}p{0.8 cm}p{1.0cm}p{0.8 cm}p{0.8 cm}}
\hline
\hline
 & \multicolumn{4}{c}{ $A = 5$} &\multicolumn{4}{c}{$A=6$} \\
  \cline{2-6} \cline{7-9} 
  & \multicolumn{2}{c}{ g.s. } &\multicolumn{2}{c}{1ex} & \multicolumn{2}{c}{ g.s. } &\multicolumn{2}{c}{1ex}\\
 \cline{2-6} \cline{7-9} \\
$K_{\max}$ & HH &  HCM & HH & HCM &  HH &  HCM & HH & HCM \\
\hline
0 & 1.9131  & 1.9131 & 0.64284 & 0.64284 & 3.7731 & 3.7731 & 2.0107 & 2.0107\\
4 & 1.9412  & 1.9366 & 0.74601 & 0.73503 & 3.8076 & 3.8024 & 2.0109 & 2.1260 \\
6 & 1.9441  & 1.9381 & 0.77879 & 0.75940 & 3.8099 & 3.8034 & 2.1662 & 2.1448 \\
8 & 1.9450  & 1.9385 & 0.80247 & 0.77666 & 3.8108 & 3.8038 & 2.1886 & 2.1606 \\
10 & 1.9452 & 1.9386 & 0.81388 & 0.78694 & 3.8109 & 3.8038 & 2.1964 & 2.1672 \\
12 & 1.9452 & 1.9386 & 0.82087 & 0.79250 & 3.8109 & 3.8038 & 2.2008 & 2.1706 \\
14 & 1.9452 & 1.9386 & 0.82484 & 0.79608&  3.8109 & 3.8038 & 2.2027 & 2.1723  \\ 
16 & & & 0.82723 & 0.79817 &  & & 2.2036 & 2.1731  \\
18 & & & 0.82867 & 0.79949 & & & 2.2040 & 2.1736  \\
20 & & & 0.82958 & 0.80032 & & & 2.2041 & 2.1738  \\
22 & & & 0.83015 & 0.80085 & & & 2.2042 & 2.1739     \\
24 & & & 0.83050 & 0.80119 & & & 2.2042 &  2.1740   \\
\hline
\hline
 & \multicolumn{4}{c}{ $A = 8$} &\multicolumn{4}{c}{$A=10$}  \\
  \cline{2-6} \cline{7-9}  
  & \multicolumn{2}{c}{ g.s. } &\multicolumn{2}{c}{1ex} & \multicolumn{2}{c}{ g.s. } &\multicolumn{2}{c}{1ex} \\
 \cline{2-6} \cline{7-9}  
$K_{\max}$ & HH &  HCM & HH & HCM &  HH &  HCM & HH & HCM   \\
\hline
0  & 9.7159 & 9.7159 & 7.1217 & 7.1217 & 18.855 & 18.854 & 15.551 & 15.551\\
4  & 9.7607 & 9.7551 & 7.2720 & 7.2568 & 18.908 & 18.902 & 15.709 & 15.695\\
6  & 9.7623 & 9.7557 & 7.2882 & 7.2685 & 18.909 & 18.902 & 15.720 & 15.703\\
8  & 9.7632 & 9.7560 & 7.3053 & 7.2794 & 18.910 & 18.902 & 15.734 & 15.711\\
10 & 9.7633 & 9.7560 & 7.3089 & 7.2822 & 18.910 & 18.902 & 15.736 & 15.712\\
12 & 9.7633 & 9.7561 & 7.3108 & 7.2834 & 18.910 & 18.902 &        & 15.713\\
14 & 9.7633 &  9.7561   &  &  7.2839 &  &   & & 15.713    \\ 
16 & & & &7.2840 &  \\
18 & & & & 7.2841 & \\
\hline
\hline
\end{tabular}      
\end{table*}

In this section, the HCM is applied   to calculate the  binding energies of helium drops with  $A=5,6,8,10$ using the same soft gaussian He-He two-body potential that was used in \cite{Gat11} and \cite{Tim12}:
\beq
V_{ij}(r) = V_0 \exp\left(-\frac{r^2}{a_0^2}\right)
\eeqn{2bpot}
with $V_0 = -1.227$ K and $a_0 = 10.03$ a.u. The $\hbar^2/m=41.281 307$ (a.u.)$^2$K value was used in the coupled system of differential equations (\ref{f5}). 
This system was solved using a modification of the  Lagrange-Laguerre mesh  method \cite{Ba15}  modified in \cite{Tim17} to deal with many-body HH methods. Only $\sim 10$ mesh points were needed to get the  ground binding energies with a five-digit precision. To achieve a similar accuracy of the energies of the first excited states,  a larger number of mesh points, about 18, is needed.

The binding energies calculated both in full HH basis (previously published in \cite{Gat11,Tim12}\footnote{The unpublished HH energies of the first excited states for $A=8$ and 10 were taken from the notes left from the 2012 study \cite{Tim12}}.) and in HCM are shown in Table 1  as functions of $K_{\max} = 2\nu_{\max} $. 
Convergence of ground state energies is very fast, with $K_{\max} \sim 10 $ needed to get five  stable significant digits. Table 1 also confirms the observation of \cite{Tim12} concerning improved convergence with increasing $A$. For excited states, the convergence of HH energy is achieved at  higher values of $K_{\max}$, between 14 and 22 depending on the number of bosons, but it also improves with $A$.  The table shows that selecting only the HCH basis states,   associated with long-range behaviour, gives a major contribution to the binding energy. The deviations of the HCM ground state energies from exact HH results are 0.3\%, 0.2\%, 0.1\%, 0.04\% for $A = 5,6,8,10$, respectively. For the first excited state they are slightly larger, being
3.7\%, 1.4\%, 0.4\%, 0.2\%.

\begin{table*}
\caption{ Root-mean-squared matter radii $r_m$ for   ground and first excited 0$^+$ states of a system of $A$ helium atoms  
obtained  for several values of $K_{\max}$ in   HCM. The r.m.s. radii $r_v$  of the overlaps  between the two states of $A$-b  and the ground state  $A-1$-body wave functions are also shown. All the radii are in a.u.
}
      \begin{tabular}{p{0.7 cm}p{0.9 cm}p{1.0 cm}p{0.8 cm}p{1.1cm}p{0.8 cm}p{1.0cm}p{0.8 cm}p{0.8 cm}}
\hline
\hline
 & \multicolumn{4}{c}{ $A=5$} &\multicolumn{4}{c}{$A=6$} \\
  \cline{2-6} \cline{7-9} 
  & \multicolumn{2}{c}{ $r_m$ } &\multicolumn{2}{c}{$r_v$} & \multicolumn{2}{c}{ $r_m$ } &\multicolumn{2}{c}{$r_v$}\\
 \cline{2-6} \cline{7-9} \\
$K_{\max}$ & g.s. &  1ex   & g.s. & 1ex & g.s. & 1ex& g.s. & 1ex\\
\hline 
    0&  6.9383&  10.950&  8.7840&  17.319&  6.3848&  8.7198&  7.7241&  14.193\\
    4&  7.0091&  11.200&  8.8400&  19.280&  6.4375&  8.9417&  7.7663&  15.773\\
    6&  7.0194&  11.388&  8.8599&  20.053&  6.4418&  9.0326&  7.7747&  16.215\\
    8&  7.0234&  11.623&  8.8639&  20.947&  6.4440&  9.1542&  7.7770&  16.746\\
   10&  7.0247&  11.859&  8.8658&  21.693&  6.4444&  9.2407&  7.7777&  17.072\\
   12&  7.0250&  12.057&  8.8662&  22.289&  6.4445&  9.3058&  7.7778&  17.305\\
   14&  7.0251&  12.242&  8.8663&  22.816&  6.4445&  9.3529&  7.7778&  17.465\\
   16&  7.0251&  12.393&  8.8664&  23.234&  6.4445&  9.3847&  7.7778&  17.571\\
   18&  7.0251&  12.523&  8.8664&  23.580&  6.4445&  9.4063&  7.7778&  17.641\\
   20&  7.0251&  12.630&  8.8664&  23.852&  6.4445&  9.4205&  7.7778&  17.686\\
   22&  7.0251&  12.717&  8.8664&  24.062&  6.4445&  9.4297&  7.7778&  17.715\\
   24&  7.0251&  12.788&  8.8664&  24.220&  6.4445&  9.4356&  7.7778&  17.733\\
   26&  7.0251&  12.844&  8.8664&  24.336&  6.4445&  9.4394&  7.7778&  17.745\\
   28&  7.0251&  12.889&  8.8664&  24.420&  6.4445&  9.4417&  7.7778&  17.752\\
   30&  7.0251&  12.924&  8.8664&  24.479&  6.4445&  9.4432&  7.7778&  17.756\\
   32&  7.0251&  12.952&  8.8664&  24.519&  6.4445&  9.4442&  7.7778&  17.759\\
   34&  7.0251&  12.974&  8.8664&  24.548&  6.4445&  9.4448&  7.7778&  17.760\\
   36&  7.0251&  12.991&  8.8664&  24.567&  6.4445&  9.4451&  7.7778&  17.761\\
   38&  7.0251&  13.004&  8.8664&  24.580&  6.4445&  9.4454&  7.7778&  17.762\\
   40&  7.0251&  13.014&  8.8664&  24.589&  6.4445&  9.4455&  7.7778&  17.762\\
   42&  7.0251&  13.021&  8.8664&  24.596&  6.4445&  9.4456&  7.7778&  17.762\\
   44&  7.0251&  13.027&  8.8664&  24.600&  6.4445&  9.4456&  7.7778&  17.762\\
   46&  7.0251&  13.031&  8.8664&  24.603&  6.4445&  9.4457&  7.7778&  17.762\\
   48&  7.0251&  13.035&  8.8664&  24.605&  6.4445&  9.4457&  7.7778&  17.763\\
   50&  7.0251&  13.037&  8.8664&  24.606&  6.4445&  9.4457&  7.7778&  17.763\\
   52&  7.0251&  13.039&  8.8664&  24.607&  6.4445&  9.4457&  7.7778&  17.763\\
   54&  7.0251&  13.041&  8.8664&  24.608&  6.4445&  9.4457&  7.7778&  17.763\\
   56&  7.0251&  13.042&  8.8664&  24.608&  6.4445&  9.4457&  7.7778&  17.763\\
   58&  7.0251&  13.043&  8.8664&  24.609&  6.4445&  9.4457&  7.7778&  17.763\\
   60&  7.0251&  13.044&  8.8664&  24.609&  6.4445&  9.4457&  7.7778&  17.763\\
      \hline
    & \multicolumn{4}{c}{ $A=8$} &\multicolumn{4}{c}{$A=10$} \\
   \hline
    0&  5.7014&  6.8970&  6.5448&  11.407&  5.2659&  6.0329&  5.8681&  10.015\\
    4&  5.7349&  7.0201&  6.5719&  12.509&  5.2898&  6.1086&  5.8876&  10.866\\
    6&  5.7362&  7.0472&  6.5746&  12.707&  5.2904&  6.1198&  5.8889&  10.980\\
    8&  5.7371&  7.0820&  6.5756&  12.923&  5.2909&  6.1332&  5.8894&  11.093\\
   10&  5.7372&  7.0975&  6.5758&  13.007&  5.2909&  6.1376&  5.8895&  11.125\\
   12&  5.7373&  7.1067&  6.5758&  13.054&  5.2910&  6.1398&  5.8895&  11.140\\
   14&  5.7373&  7.1112&  6.5758&  13.076&  5.2910&  6.1406&  5.8895&  11.146\\
   16&  5.7373&  7.1134&  6.5758&  13.086&  5.2910&  6.1409&  5.8895&  11.148\\
   18&  5.7373&  7.1145&  6.5758&  13.091&  5.2910&  6.1411&  5.8895&  11.149\\
   20&  5.7373&  7.1150&  6.5758&  13.093&  5.2910&  6.1411&  5.8895&  11.149\\
   22&  5.7373&  7.1152&  6.5758&  13.094&  5.2910&  6.1412&  5.8895&  11.149\\
   24&  5.7373&  7.1154&  6.5758&  13.095&  5.2910&  6.1412&  5.8895&  11.149\\
\hline
\hline
\end{tabular}      
\end{table*}

The r.m.s. radii are calculated from 
\beq
\la r^2 \ra &=& \frac{1}{A} \sum_{i=1}^A (\ve{r}_i - R_{c.m.}) = \frac{1}{A} \la \rho^2 \ra \\
\la \rho^2 \ra  &=&   \sum_{\nu} \int_0^{\infty} d\rho \, \rho^2 \chi^2_{\nu}(\rho) 
\eeqn{r2}
on the condition that
$\sum_{\nu} \int_0^{\infty} d\rho \, \chi^2_{\nu}(\rho)=1$, which was always verified to be true in the calculations.
The r.m.s. radii $r_m = \sqrt{\la r^2 \ra }$, obtained in HCM, are collected in Table II. For ground states, to achieve converged  $r_m$ values   up to five significant digits  $K_{\max}= 14$ and 12 are needed for $A = 5$ and 6-10, respectively. For excited states, the $r_m$ convergence for same number of significant digits is achieved  with  $K_{\max} = 60, 46, 24, 22$. With increasing $A$ the r.m.s. radii decrease both for ground and the first excited states, as expected for bosonic systems interacting via a soft two-body potential and undergoing a collapse. This decrease is more significant for excited states, which is related to increasing one-boson separation energies with $A$, shown in Table III.

From equations (\ref{f1}) and (\ref{r2}) one can conclude that the average r.m.s. distance between two boson in a drop is $r_{ij} = \sqrt{\frac{2A}{A-1}} \,r_m$.  Therefore, any two bosons  are separated by 11, 10, 8.7 and 7.9 a.u. in   the ground state of drops with $A=5,6,8,10$, respectively. 

\section {Overlap Integrals}
\label{sec:6}

\subsection{Formal aspects}
The most interesting application of the HCM is in projecting  the many-body wave function $\Psi_A$ of  $A$ particles onto the wave function $\Psi_{A-1}$ of the $A-1$-body system. 
Such projections, called  the overlap integrals in nuclear physics and Dyson orbitals in atomic and molecular physics, are defined as 
\beq
I(\ve{r}) = \sqrt{A} \la \Psi_{A-1} (1,...,A-1) | \Psi_{A}(1,...,A-1,A), \ra
\eeqn{overlap}
where $\Psi_{A-1}$ and $\Psi_A$ are translation-invariant and the integration is performed over the $A-2$ internal Jacobi coordinates of $A-1$. These Jacobi coordinates are not normalised (which means there is no $\sqrt{i/(i+1)}$ factor in the definition (\ref{Jacobi})) so that $\ve{r}$ is the physical distance between the centre of mass of $A-1$ and the particle $A$ that does not belong to $A-1$. The $\sqrt{A}$ factor is added to the definition to take into account the need for (anti)symmetrisation between identical particles in reaction calculations.

The most interesting properties of the overlap integrals discussed in nuclear physics are spectroscopic factors and asymptotic normalization coefficients (ANCs). The spectroscopic factors 
\beq
S_l = \int_0^{\infty} dr\, r^2 I^2_l(r),
\eeqn{sf}
are defined as the norms of the radial parts $I_l(r)$ of $I(\ve{r}) $ 
corresponding to the orbital momentum $l$. The spectroscopic factor  can be related to occupancies of the single-particle orbitals of the many-body system \cite{Mac60}. It can also be interpreted as the numbers of $(A-1)$ systems in a chosen state   inside $A$. An example of such interpretation can be found in \cite{Sch86}. In atomic physics  $S_l$ are related with the probability of a selected ionization channel and are often called probability factors or pole strengths \cite{Ort20}.

The $I_l(r)$ satisfies the Schr\"odinger-like inhomogeneous equation with the source term determined by $\la \Psi_{A-1} | \sum_{i \in (A-1)} V_{iA} | \Psi_A\ra$ \cite{Tim14}. For short-range two-body potentials this source term vanishes at large $r$, which leads to an exponential decrease of $I_l(r)$ with the decay constant of $\kappa = \sqrt{2m\mu (E_{A-1}-E_A)}/\hbar$, where $\mu = (A-1)/A$ is the reduced mass number for  the core plus one boson system, determined by the difference of (negative) binding energies $E_{A-1}$ and $E_A$ of $A-1$ and $A$. For $l=0$ this asymptotic behaviour is given by
\beq
I_{as}(r) = C \, \frac{\exp(-\kappa r)}{r},
\eeqn{asymp}
where $C$ is the ANC. It is related to the probability amplitude for the last boson to stay in a classically-forbidden region and it is also  connected by a simple renormalization factor to a vertex constant (or on-shell amplitude) for $A\rightarrow (A-1)+1$ decay - the quantity traditionally used in particle physics - and both the ANC and the vertex constant are very sensitive to the choice of the two-body interaction used to calculate them \cite{Bay92}. 
Reproducing the correct form of $I(\ve{r})$ in microscopic calculations is of paramount importance for predicting the cross sections of reactions sensitive to the region of space where one particle is far away from the rest. Nuclear physics deals with many reactions of this type, which are often needed for astrophysical applications \cite{Muk22}. The HH and HCH bases are well suited well for such a purpose. Convergence of $I(r)$ with increasing HH model space has been demonstrated in \cite{Tim07,Tim08b}. Open problems in overlap integral calculations are reviewed in \cite{Tim14}.

In constructing the  overlap integrals $\la A-1 \mid A\ra$ the
  $\Psi_{A-1}$ wave function was  taken at   the lowest order of the HH expansion,  $K=0$, to be consistent with the description of the $A-1$ core of $A$. Its hyperradial part is given by one function only, $\chi^{(A-1)}_0(\rho_c)$. The   $\Psi_A$ was taken from the HCM and its radial parts are labelled as $\chi^{(A)}_{\nu}(\rho)$. With these assumptions,
following Ref. \cite{Tim07}, one can easily derive the  expression 
for    the radial part $I(r)$ of this overlap, defined via $I(\ve{r}) = I(r) Y_{00}(\hat{r})$, as 
\beq
& &I (r)
= \mu^{3/4} 
  \sum_{\nu=0}^{\nu_{\max}}N^{(n)}_{0 \nu 0} \sqrt{{\cal N}_{\nu}} 
  \eol
 & \times&
  \int_0^{\infty} d \rho_c \,
 \frac{\rho_c^{\frac{n-4}{2}} \chi_0^{(A-1)}(\rho_c)\, \chi^{(A)}_{\nu}
  \left( \sqrt{\rho_c^2+\mu r^2}\right)
 }
 {\left(\rho_c^2+\mu r^2\right)^{\frac{n-1}{2}}}
P_{\nu}^{\frac{n-5}{2},\frac{1}{2}}\left( \frac{\mu r^2-\rho_c^2}{\mu r^2+\rho_c^2}\right).\,\,\,\,\,\,\,\,\,\,\,\,\,\,
  \eeqn{Ir}
  Here the $\mu^{3/4}$ factor arises because   the hyperradii and hyperangles in HH and HCM    are built on normalized Jacobi coordinates while the overlaps are usually the  functions of the physical distance between the removed particle and the core.

\begin{figure}[t]
\vspace{0.5 cm}
\includegraphics[scale=0.24]{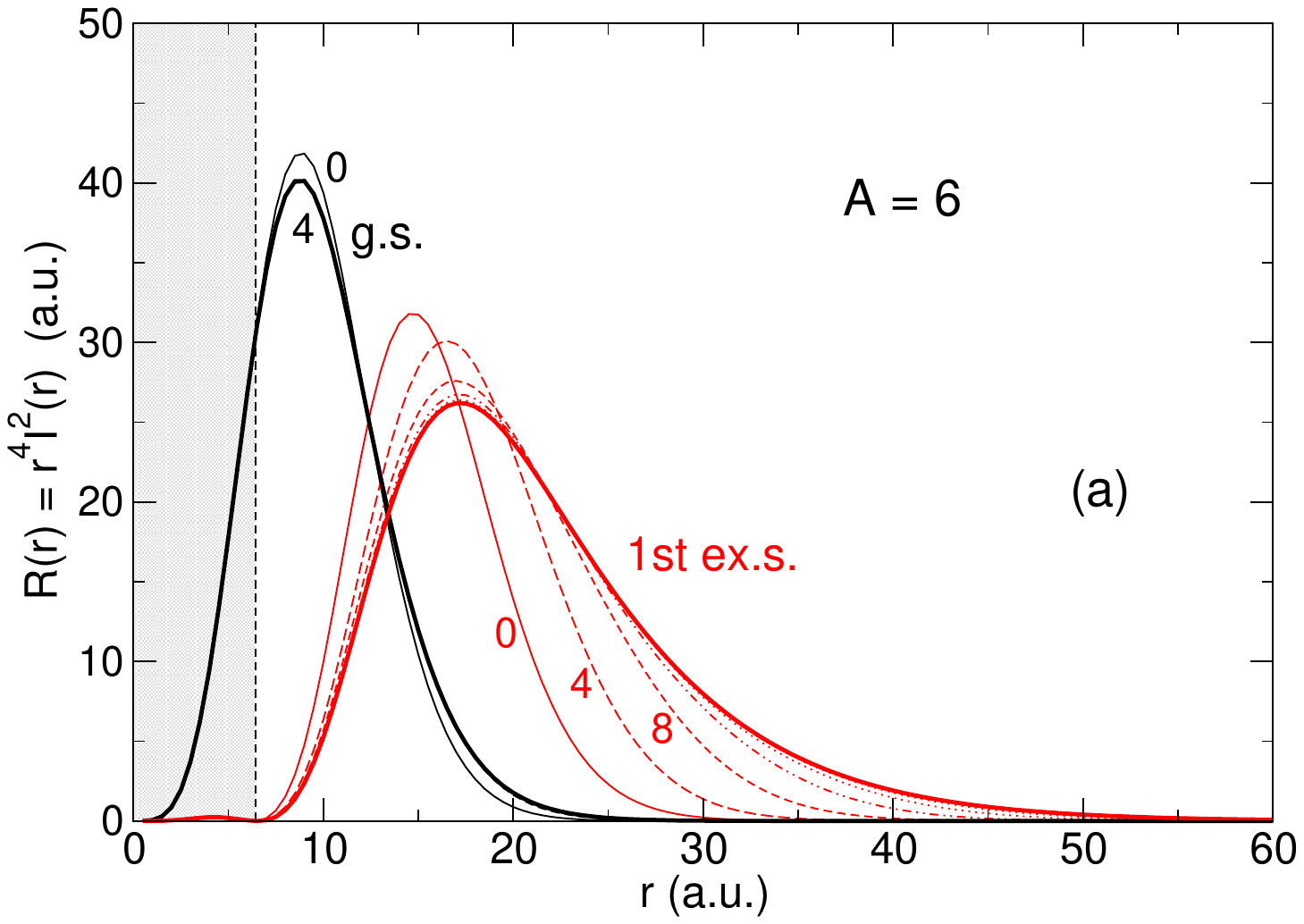}
\includegraphics[scale=0.24]{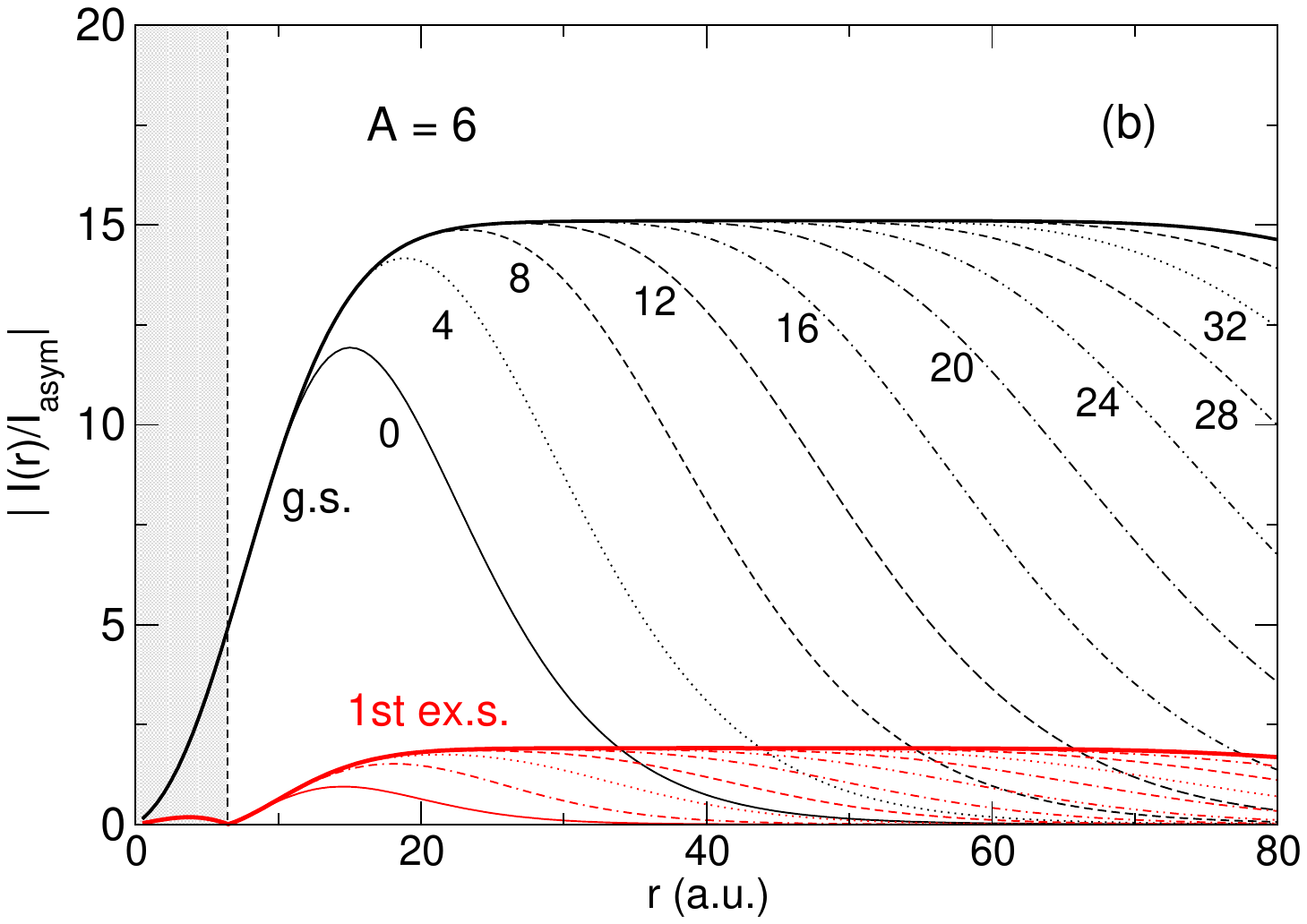}
\caption{($a$) The $R(r) = r^4I^2(r)$ function and ($b$) the absolute values of $I(r)/I_{as}(r)$ shown for $K_{\max}$ from 0 to 40 with a step of 4 for the ground and the first excited states of $A=6$. Grey area corresponds to $r < r_m({\rm g.s.}) $. }
\label{fig:asym}
\end{figure}

\subsection{Numerical calculations: overlaps and their radii}
 
  When performing numerical calculations two type of convergence were considered: convergence of $I(r)$ with increasing number of Lagrange-Laguerre mesh points at a fixed $\nu_{\max}$ and its convergence with increasing $\nu_{\max}$. A small number of mesh points (between 10 and 20) is sufficient for getting accurate values of $I(r)$ at small $r$ only. To get a convergence at large distances $r \sim 60-80$ a.u. up to 50 mesh points were needed. Therefore, all overlap calculations presented below were obtained with 50 mesh points. A typical example of  convergence  
  with $\nu_{\max}$  is shown in Fig. \ref{fig:asym}a for the quantity $R(r)=r^4I^2(r)$ for $A=6$. This function carries   information about the contribution to the r.m.s. radius $r_v = \left(  \int_0^{\infty} dr R(r)/\int_0^{\infty} dr S(r)\right)^{1/2}$ of $I(r)$, where $S(r) = r^2 I^2(r)$, which could be associated with the radius  of the last (or valence) particle orbit. The $r_v$ values are very close to the position of the maxima in $R(r)$. They are shown in Table 1 as a function of $K_{\max} = 2\nu_{\max}$.  For the ground state the convergence of $R(r)$ and $r_v$ is achieved very fast but a much larger model space is needed for excited states. This is related to a significant contribution to $r_v$ from $I(r)$ and $R(r)$ at large $r$, up to 50 a.u., which requires   an accurate asymptotic description of these functions in that area. The convergence of $I(r)/I_{as}(r)$ is shown in Fig. \ref{fig:asym}b for $A=6$. With increasing model space it converges to a constant (the ANC $C$) after 22 a.u.  For other helium drops
  the convergence of $I(r)$, $R(r)$, $r_v$  and $C$ (not shown here) is similar to the case of $A=6$ and the asymptotic behaviour of $I(r)$ is achieved approximately in the same area between 20 and 25 a.u. The values of ANCs, together with  spectroscopic factors $S$ are given in Table 3 for all the helium drops considered here. 
  \begin{table*}[t]
\begin{center}
\caption{ Separation energies $\varepsilon = |E_A - E_{A-1}(K_c=0)|$ (in K), spectroscopic factors $S$ and asymptotic normalization coefficients $C$ (in a.u.$^{-1/2}$) calculated for ground and first excited states in a system of $A$ helium atoms. 
}
      \begin{tabular}{p{0.8 cm}p{1.0 cm}p{1.0 cm}p{1.8 cm}p{1.0 cm}p{1.0cm}p{ 1.1 cm}}
\hline
\hline
&   \multicolumn{3}{c}{ g.s.} &\multicolumn{3}{c}{1ex} \\
  \cline{2-4} \cline{5-7} 
$A$ & $\varepsilon$ & $S$  & $C$ &  $\varepsilon$ & $S$ & $C$ \\
\hline
5  & 1.213 &  4.6666 &    7.9618 &   0.0752 &  1.3608 &        -0.80711 \\
6  & 1.891  &  5.7356 &  15.109 &  0.2288 &  1.3206 &          -1.9075  \\
8  & 3.396  &  41.733 & 7.7788 &  0.9791 &   1.2501 &           -6.8833 \\
10 & 5.030  &  9.7892 & 94.365 &  1.8414 &   1.2073 &          -18.051\\
\hline
\hline
\end{tabular}    
\end{center}
\label{SFandANC}
\end{table*}

\subsection{Classically-forbidden region and halo}

Fig. \ref{fig:asym}a shows that in the shallow excited state the last (or valence) atom stays far outside the r.m.s.  $r_m$ of the drop, forming a two-body molecule. On this grounds such a state  could be called "halo". However, the concept of "halo" is often associated with a significant probability of finding the valence particle in the classically-forbidden region. Examining the overlaps $I(r)$  and their contribution $S(r)$ to the spectroscopic factor in different regions of $r$ can help the understanding  halo formation in shallow excited states.
    
\begin{figure}[t]
\vspace{0.5 cm}
\includegraphics[scale=0.24]{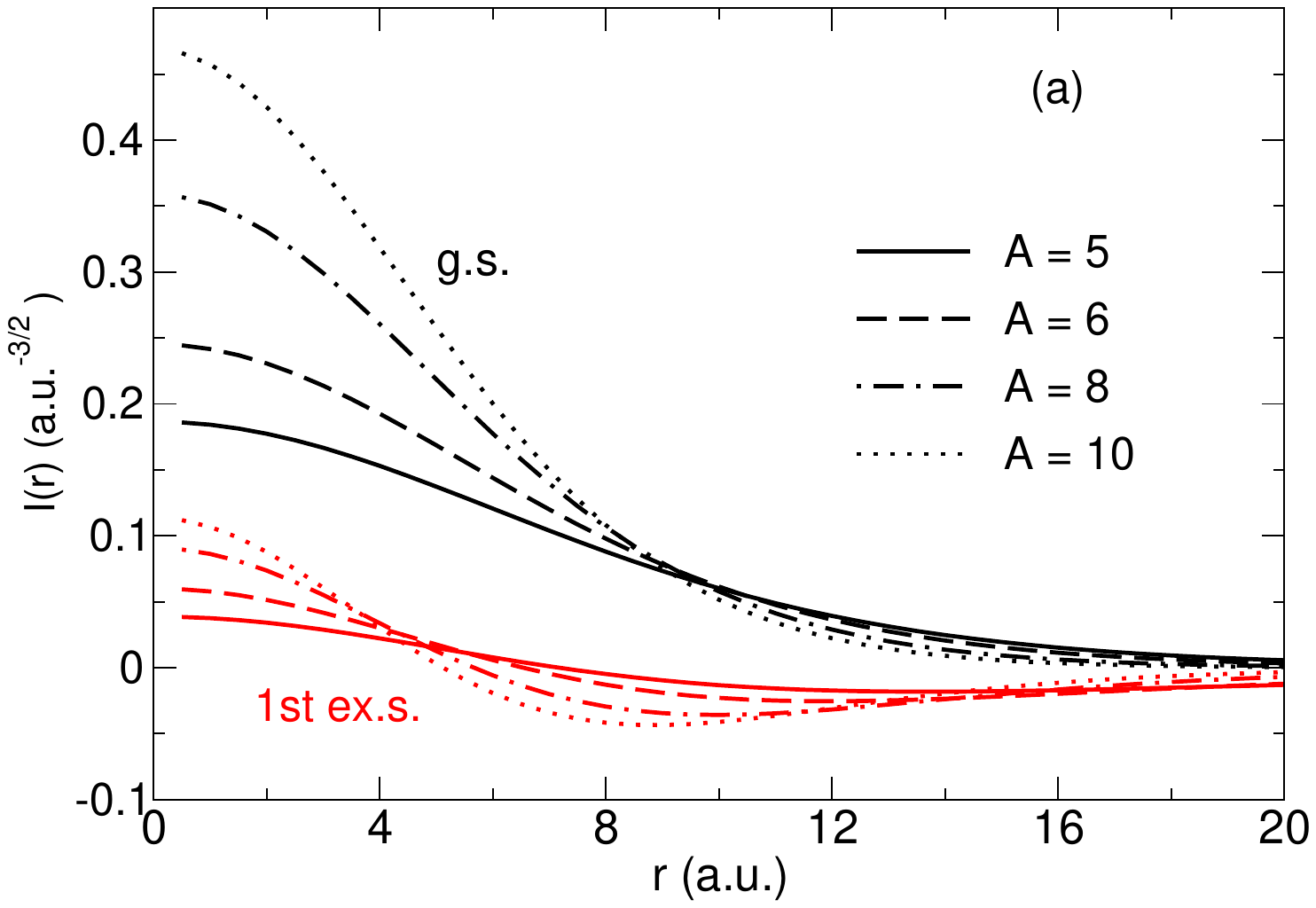}
\includegraphics[scale=0.24]{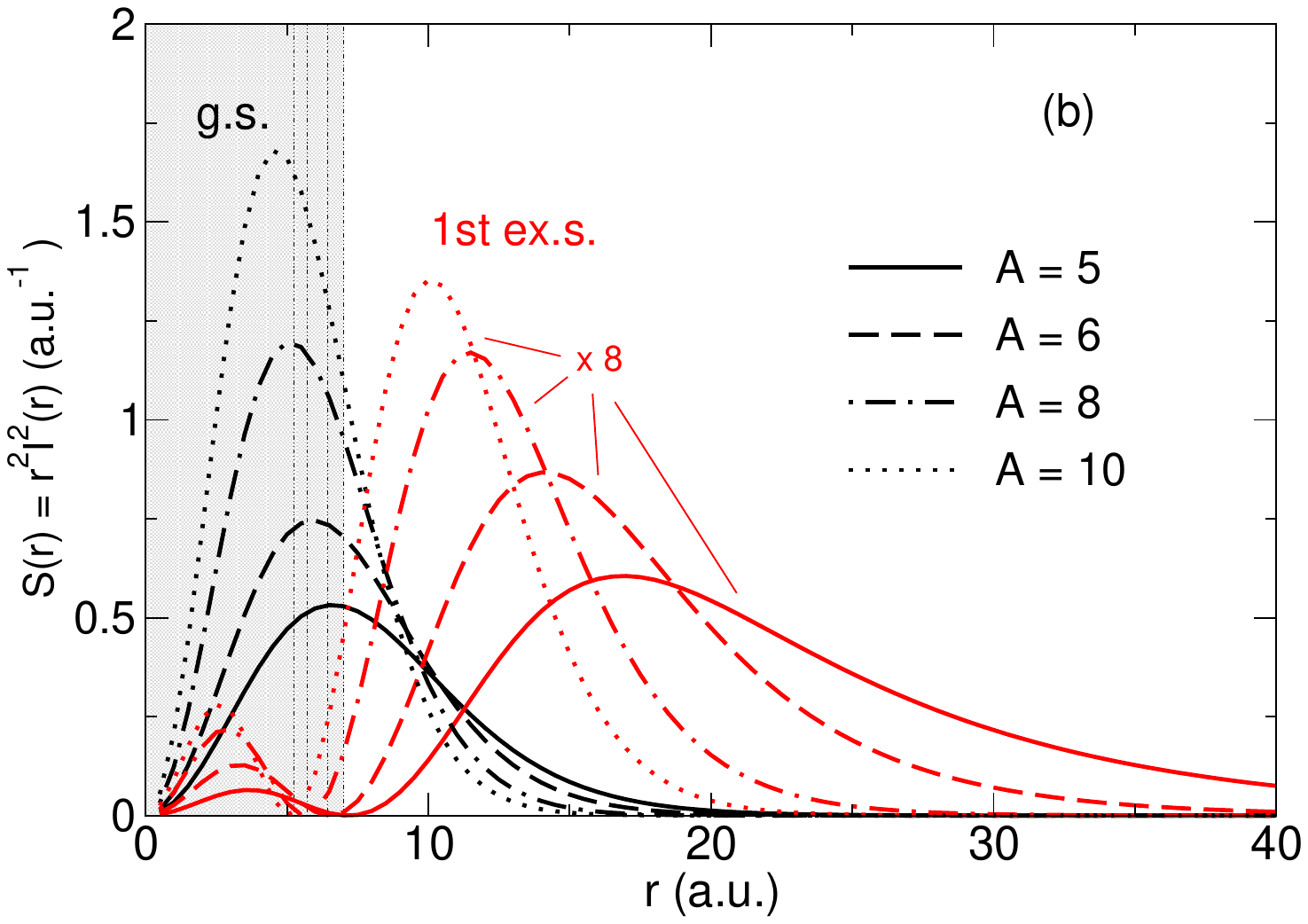}
\caption{Overlap integrals $I(r)$ ($a$)  and  the corresponding $S(r)=r^2I(r)^2$ values ($b$) calculated for ground and first excited stats of  5, 6, and 10 helium atoms. Grey area corresponds to $r < r_m$ for $A=5$ in ground state and other grey vertical lines show $r_m$ values for six, eight and ten atoms (from right to left).}
\label{fig:overlaps}
\end{figure}

The converged   overlaps of $\Psi_A$ with the ground state wave function $\Psi_{A-1}$ of $A-1$ are shown in Fig. \ref{fig:overlaps}a for $A=5, 6, 8, 10$. For ground states of $A$ they have a shape typical of $0s$  single-particle wave functions while for the first excited  states they have one node, resembling   $1s$ single-particle wave functions, as expected. The quantity $S(r)$  
shown in Fig. \ref{fig:overlaps}b, gives an idea where a single helium atom  could be located within the drop.  For ground states the maximum of $S(r)$ is located around the r.m.s. radius of the drop $A$, which is about 7, 6.4, 5.7 and 5.3 a.u. for five, six, eight and ten atoms, respectively. In all cases $S(r)$ are negligible beyond 20 a.u. For a boson located there  the   strength of its interaction with an atom located at the drop's surface, given by  (\ref{2bpot}), is about 20-30\% of that experienced by two bosons located inside the drop at an average distance $r_{ij}$. Therefore, it is fair to say that all atoms in the ground states of helium drops are confined within the range of their interaction.
 
The maxima of $S(r)$  in  the   shallow excited states are shifted well outside $r_m$ indicating a significant probability to find the last atom far away from the $A-1$ core (see  Fig. \ref{fig:overlaps}b). Whether this region can be associated with the  classically-forbidden one depends on  the  definition of the latter. There are two ways of thinking about such a definition. 
\begin{itemize}
\item  Classically-forbidden region $r_{\rm c.f.}$ could be defined as the region of $r$ where $I(r)$ is determined by its asymptotic form $I_{as}(r)$ given by (\ref{asymp}). In this region a trivially-equivalent local effective potential,   generating $I(r)$ from a two-body Schr\"odinger equation, vanishes. Fig. \ref{fig:asym}b shows that for $A=6$ such a region starts somewhere after 20 a.u. for both the ground and excited states, well beyond the radius $r_m$ of the helium drop. In the intermediate region $r_m < r < 20$ a.u.     the He-He interaction can still influence the $I(r)$ behaviour because its range is 10.03 a.u. A similar situation occurs for all other systems considered, with $r_{\rm c.f.} \sim 20$-25 a.u. With such a definition of $r_{\rm c.f.}$ the probability that a  loosely-bound boson   stays iwithn the classically-forbidden region is not very big, even for $A=5$ so that such states cannot be called ``halo" despite showing a well-developed molecular structure and being located far outside $r_m$. 
\item
In two-body  problem the classically forbidden region is associated with the region of space where the kinetic energy of the particle is negative. In a similar way, one can accept  the region where $-\frac{\hbar^2}{2\mu m} \frac{\partial^2}{\partial r^2} I(r) < 0 $ as the definition of $r_{\rm c.f.}$. For excited states of  $A=5,6,8,10$ this value is 19, 16, 13 and 12 a.u., respectively, and there is a significant probability to find the last boson in the area beyond $r_{\rm c.f.}$. The main contribution to the r.m.s. radii $r_v$ also comes from this area. However, for ground states all $r_{\rm c.f.}$ are around 4-5 a.u. and with this definition even the tighly-bound ground states would have a significant probability for finding the last boson  in classically-forbidden range despite $r_{\rm c.f.}$ being much smaller than the He-He interaction range.
\end{itemize}
Irrespective of  the definition, the probability of one atom staying far way from the rest depends on its separation energy. With the chosen He-He interaction this separation energy increases with $A$ (see Table 2) decreasing the size of the two-body molecule in the shallow excited state. With a further increase of $A$ one can expect that the valence atom in the first excited state would be located mainly in the range of the He-He interaction, further demonstrating a collapse of bosonic systems interacting with a soft two-body force only. Including repulsive three-body forces prevents the collapse \cite{Kie17} but this task  is beyond the scope of the present study.

\subsection{Spectroscopic factors and sum rules}

 \begin{table*}[t]
\begin{center}
\caption{  Spectroscopic factors $S$ calculated for ground and first excited states in a system of $A$ helium atoms for overlaps with sevelap excited states of $A-1$. 
}
      \begin{tabular}{p{0.9 cm}p{0.7 cm}p{1.0 cm}p{0.7 cm}p{1.0 cm}p{0.7 cm}p{1.0 cm}p{0.7 cm}p{0.6cm}}
\hline
\hline
&   \multicolumn{2}{c}{ $A=5$} &\multicolumn{2}{c}{$A=6$}&\multicolumn{2}{c}{$A=8$}&\multicolumn{2}{c}{$A=10$} \\
  \cline{2-9} 
$A-1$ &   g.s.  &  1ex & g.s. & 1ex &   g.s.  &  1ex & g.s. & 1ex \\
\hline
g.s. & 4.67 & 1.36 & 5.74 & 1.32 & 7.78 & 1.25 & 9.79 & 1.21 \\
1ex  & 0.11 & 1.39 & 0.17 & 3.31 & 0.16 & 5.85 & 0.14 & 8.03 \\
2ex  &      & 0.03 & 0.02 & 0.25 & 0.01 & 0.30 & 0.01 & 0.27 \\
3ex  & 0.01 & 0.09 &      & 0.01 &      & 0.04 &      & 0.02 \\
4ex  & 0.01 & 0.13 &      & 0.02 &      & 0.01 \\
5ex  & 0.02 & 0.15 &      & 0.02 \\
6ex  & 0.02 & 0.15 &      & 0.03 \\
7ex  & 0.02 & 0.13 &      & 0.03 \\
8ex  & 0.02 & 0.12 &      & 0.03 \\
9ex  & 0.02 & 0.10 &      & 0.03 \\
10ex & 0.02 & 0.08 &      & 0.03 \\
11ex & 0.01 & 0.07 &      & 0.03 \\
12ex & 0.01 & 0.05 &      & 0.03 \\
13ex & 0.01 & 0.04 &      & 0.02 \\
14ex & 0.01 & 0.03 &      & 0.02 \\
15ex & 0.01 & 0.02 &      & 0.02 \\
\hline
sum  & 4.97 & 3.94 & 5.93 & 5.21 & 7.95 & 7.45 & 9.94 & 9.53 \\
max & 5    & 5 & 6 & 6 & 8 & 8 & 10 & 10 \\
\hline
\hline
\end{tabular}    
\end{center}
\end{table*}

Figure \ref{fig:overlaps}a shows that the  absolute values of the overlaps increase with $A$ and that they are much larger for ground states  than those for excited states, which is reflected in spectroscopic factors shown in Table 4. For a harmonic oscillator two-body interaction the spectroscopic factors between the ground states would be equal to $A$. 
For any other interaction the spectroscopic factors are smaller than that because of the non-zero overlap  of the wave function of  $A$ with that of any state of $A-1$. The spectroscopic factors satisfy the sum rules
$\sum_{i} S_i^{(m)}=A$, where $S_i^{(m)}$ is associated with the overlap of the $m$-th state of $A$ with $i$-th state of $A-1$. The sum rule follows from the orthonormality of $\Psi_A$ and including the factor $A$ into the definition of spectroscopic factors. 

For the soft potential used here, the $S^{(g.s.)}_{g.s.}$ values are close to the maximum allowed value   approaching to it with increasing $A$. It almost exhausts the sum rules for ten atoms. The remaining contribution comes from the overlap of the ground state of $A$ with excited states of $A-1$, shown in  Table 4  only for those  larger than 0.01. Adding only one excited shallow bound state gives 95\%, 98\%, 99.2\% and 99.3\% contribution to the sum rules for $A=5,6,8,10$ atoms, respectively. The rest comes from the  excited states of $A-1$ above the breakup threshold, which is shown for $A=5$ only. This contribution is broadly distributed between the 3rd and 15th excited states in the positive energy range between 0.1 and 5 K. 

The situation is very different for overlaps of the shallow excited state of $A$ with the ground state of $A-1$. The $S^{1ex}_{g.s.}$ values are between 1.36 and 1.21, which could be interpreted as in this state there is  approximately one partition corresponding to the molecule formed by the $A-1$ ground state and the $1s$ valence atom. A closer examination reveals that the main contribution to the sum rule for the shallow state comes from its overlap with the shallow excited state of $A-1$ (see Table 4). This is because each atom of the   $A-1$ core of the shallow molecular excited state of $A$ 
finds itself in the $0s$ state with respect to the other $A-1$ atoms formed by the remaining $A-2$   core atoms and the valence $1s$ atom  of $A$. Given that $A-1$ atoms could be selected from the core of the  excited state of $A$, the number of partitions ``excited state of $A-1$ plus a tightly-bound $0s$ atom" should be close to $A-1$, which is reflected in Table 4. 
 The  large $S^{1ex}_{1ex}$ values could lead to larger cross sections for  removing an atom from the core of the excited state rather than for removing its valence atom.  In nuclear physics this has been seen experimentally in $(d,p)$ reaction of the halo nucleus $^{11}$Be, composed by the $^{10}$Be core and a valence neutron, when leaving the core $^{10}$Be in excited weakly-bound state has larger cross section \cite{Win01}.
The contribution to the sum rules from excited states of $A-1$ above the breakup threshold increases with the decreasing binding energy of shallow states.  It is small for eight and ten atoms but for five and six atoms 15 excited states take you only to 78\% and 86\% of the sum rules, respectively (see Table 4).

\section{Summary and conclusions}
\label{sec:7}

To describe long-range behaviour of one particle in a few- or a many-body system a model has been developed that uses expansion on HCH basis. The structure of these basis states is similar to the one used in microscopic cluster models for identical particles. It is represented by a symmetrized product of the HH of the core and a set of Jacobi-polynomial-based hyperangular functions describing relative motion of the valence  particle. 
The size of the model space, determined by the number of nodes of Jacobi polynomials and  needed  for converged results, depends on the nature of the quantities to be calculated and on the two-body interaction chosen.

A soft gaussian potential has been employed to describe light helium drops with up to ten atoms. Comparison with calculations employing a full HH basis has shown that HCM gives a major contribution to the binding energies of the drops.  While the ground state binding energies converge very fast,   more HCHs are needed for shallow first excited states, especially for lighter drops with five and six atoms. A similar convergence pattern appears for r.m.s. radii: a fast convergence for ground states and a slower convergence for first excited states.

The overlaps of the wave functions of two drops differing by one atom (or Dyson's orbitals), calculated in HCM, reveal that this atom is most likely  located at the surface of the drop in the ground state. 
It stays far outside this surface in the first excited shallow state,  forming a two-body molecule  but still being within the range of the He-He interaction. The size of the molecule, if defined as the r.m.s. of the overlap, decreases with $A$ because of increasing one-atom separation energy.  The probability of this atom to stay in the region of space where the overlap takes its asymptotic form is significant for five and six atoms but becomes small for eight and ten atoms. For the last two cases, the ground state spectroscopic factors  are close to the sum rule values  while for five and six atoms a significant contribution to the sum rules comes from excited core states. The shallow excited states overlap strongly with shallow states of the core while their overlaps with the core ground states have spectroscopic factors of the order one.  For five and six atoms a significant contribution to the sum rules comes from four-and five-body excited states above the one-atom removal threshold.

The present work used the HHs of the simplest  structure for the core so that the resulting overlaps, while giving a correct qualitative picture, could not be sufficiently accurate for their further applications. Also, these results should  strongly depend on the choice of  Hamiltonian. It is known that to avoid   collapse of  bosonic systems, such as helium drops, with increasing number of atoms a three-body interaction could be introduced. Future work will further develop the HCM to address these issues.

Finally, the version of the HCM described above can be generalized to a system of identical fermions. Given that for intrinsic  spin $J$ only $2J+1$ fermions could be represented by the HCH with $K_c=0$, it is crucial to develop  coordinate transformaitons for arbitrary products of hyperangular functions.
Availability of  fractional parentage coefficients for arbitrary angular momentum is also vital. These coefficients could be taken, for example, from \cite{Bar97}.


\renewcommand{\thesubsection}{\Alph{subsection}}
\renewcommand{\theequation}{\Alph{subsection}.\arabic{equation}}
\setcounter{equation}{0}

\section*{Appendix}

\subsection{Transformation coefficients}
\label{sec:A}

The HCM developed here relies heavily on using the transformation, given by eq. (\ref{TC}), to calculate various matrix elements. For a three-body problem such transformations are given by  Raynal-Revai coefficients \cite{RR}. For a many-body problem a similar transformation was mentioned in \cite{SSh} and it is also a part of developments of HH techniques in  \cite{Viv98,Doh20}. 

With the  current choice of $K_c=0$, only selected transformation coefficients are needed.
 The first coefficient, $T^{(n)}_{0,000}(\gamma_1,\gamma_2)$, is equal to one. All other coefficients $T^{n}_{\nu+1,\nu_1 \nu_2 l}(\gamma_1,\gamma_2)$   are then obtained from $T^{n}_{\nu,\nu'_1 \nu'_2 l'}(\gamma_1,\gamma_2)$, which is shown below. 
 
 The starting point for this derivation is
the recurrence relations of hyperangular functions
$\varphi^{(n)}_{K\nu lm}  $, which could be easily obtained from recurrence relations of Jacobi polynomials \cite{AS65}
\beq
\varphi^{(n)}_{K\nu+1 lm} 
(\theta,\hat{\xi}) &=&
\frac{{\cal N}^{(n)}_{K\nu+1 l}}{{\cal N}^{(n)}_{K\nu l}} a^{(n)}_{K \nu l} \varphi^{(n)}_{K \nu lm } (\theta,\hat{\xi})
+
\frac{{\cal N}^{(n)}_{K\nu+1 l}}{{\cal N}^{(n)}_{K\nu l}} b^{(n)}_{K\nu l} \cos 2\theta \varphi^{(n)}_{K\nu lm} (\theta,\hat{\xi})
\eol
&-&
\frac{{\cal N}^{(n)}_{k\nu+1 l}}{{\cal N}^{(n)}_{K\nu-1 l}} c^{(n)}_{K\nu l} \varphi^{(n)}_{K\nu-1 lm} (\theta,\hat{\xi}),
\eeqn{recrelphi}
where the coefficient $a,b,c$ are given by 
\beq
a^{(n)}_{K\nu l}& =&\frac{(2\nu+n/2+l-1)(n/2-l-3)(n/2+l-2)}{2(\nu+1)(\nu+n/2+l-1)(2\nu+n/2+l-2)}, 
\eol
b^{(n)}_{K\nu l} &=&
\frac{(2\nu + K+n/2 + l )(2\nu + K+n/2 + l-1) }{(\nu+1) (n+ 2\nu+2K   + 2l-2)},
\eol
c^{(n)}_{K\nu l} 
&=& \frac{ (2\nu+ n-5 )(\nu+l+1/2)(2\nu+n/2+l)}{2(\nu+1)(\nu+n/2+l-1)(2\nu+n/2+l-2)}.
\eeqn{bcoef}
The transformation coefficient $T^{(n)}_{\nu+1,\nu_1 \nu_2 l}(\gamma_1,\gamma_2)$ is defined as an overlap of the product $\varphi^{(n-3)}_{000} (\theta_2,\hat{\xi}_2)  \varphi^{(n)}_{0\nu+1 0} (\theta_1,\hat{\xi}_1)$, expressed in coordinates $\{\theta_2,\hat{\xi}_2,\theta_1,\hat{\xi}_1\}$, with the product
$[\varphi^{(n-3)}_{0\nu_2 l} (\theta'_2,\hat{\xi}'_2) \times \varphi^{(n)}_{K'\nu_1 l} 
(\theta'_1,\hat{\xi}'_1)]_{0}$ of the hyperangular functions  in new coordinates 
$\{ \theta'_2,\hat{\xi}'_2,\theta'_1,\hat{\xi}'_1\}$, related to the original ones by (\ref{c-trans}). Using (\ref{recrelphi}) for 
$\varphi^{(n)}_{0\nu+1 0} (\theta_1,\hat{\xi}_1)$, the orthogonality relation
\beq
\la \varphi^{(n)}_{K\nu l} (\theta,\hat{\xi})
\mid 
   \varphi_{K\nu'l}  ^{(n)}
(\theta,\hat{\xi})\ra = \delta_{\nu\nu'}
\eeqn{ortho}
and re-expressing $\varphi^{(n)}_{0\nu+1 0} (\theta_1,\hat{\xi}_1)$  in new coordinates with the help of (\ref{TC}),
one can obtain
\beq
& &T^{(n)}_{\nu+1,\nu_1 \nu_2 l}(\gamma_1,\gamma_2) = 
\frac{N^{(n)}_{0\nu+10}}{N^{(n)}_{0\nu 0}}b^{(n)}_{0\nu0}\sum_{ \nu'_1+ \nu'_2+ l''=\nu} T^{(n)}_{\nu, \nu'_1 \nu'_2 l''} (\gamma_1,\gamma_2)\,
\eol
&\times&
\la [\varphi^{(n-3)}_{0\nu_2 l} (\theta'_2,\hat{\xi}'_2) \times \varphi^{(n)}_{K'\nu_1 l} (\theta'_1,\hat{\xi}'_1)]_{0}
\mid 
 \left[\varphi_{0\nu'_2l''} ^{(n-3)} 
(\theta'_2,\hat{\ve{\xi}}'_{2})\times
\cos 2\theta_1   \varphi_{K''\nu'_1l''}  ^{(n)}
(\theta'_1,\hat{\xi}'_{1})\right]_{0}\ra,
\eol
\eeqn{}
in which $K'' = 2\nu'_2+l''$. Next,  using 
\beq
\cos 2\theta_1&=& 2\cos^2 \theta_1-1 = 2\,\,\frac{(\gamma_1 \vec{\xi}'_1 + \gamma_2 \vec{\xi}'_2)^2}{\rho^2}-1
\eol
&=& 2 \gamma_1^2 \cos^2 \theta'_1 +4 \gamma_1 \gamma_2 \cos \theta'_1 \sin \theta'_1 \cos \theta'_2 \,(\hat{\vec{\xi}}_1 \cdot \hat{\vec{\xi}}_2)+ 2 \gamma_2^2 \cos^2 \theta'_2 \sin^2 \theta'_1 - 1
\eol
\eeqn{cos2theta}
where $\hat{\vec{\xi}}$ is the unit vector in the direction of $\vec{\xi}$, the transformation coefficient is reduced to the following four terms,
\beq
T^{(n)}_{\nu+1,\nu_1 \nu_2 l}(\gamma_1,\gamma_2)
= (T.1) + (T.2) + (T.3) + (T.4),
\eeqn{Tsum}
that correspond to four different terms in $\cos 2\theta_1$. The last one of them, $(T.4)$, is zero because it involves the matrix element $\la [\varphi^{(n-3)}_{0\nu_2 l} (\theta'_2,\hat{\xi}'_2) \times \varphi^{(n)}_{K'\nu_1 l} (\theta'_1,\hat{\xi}'_1)]_{0}
\mid 
 \left[\varphi_{0\nu'_2l''} ^{(n-3)} 
(\theta'_2,\hat{\ve{\xi}}'_{2})\times
   \varphi_{K''\nu'_1l''}  ^{(n)}
(\theta'_1,\hat{\xi}'_{1})\right]_{0}\ra$ in which 
$ \nu_1+\nu_2+l' = \nu+1 \neq \nu = \nu'_2 + \nu'_1 + l''$. The expressions for the other three terms are derived below.

\subsubsection{Contribution from (T.1)}

This term comes from  $ 2 \gamma_1^2 \cos^2 \theta'_1$ and it is given by the following expression,
\beq
(T.1) &=&\gamma_1^2\frac{N^{(n)}_{0\nu+1 0}}{N^{(n)}_{0\nu 0}} b^{(n)}_{0\nu0}
\sum_{ \nu'_1 +\nu'_2+ l''=\nu} T^{(n)}_{\nu, \nu'_1 \nu'_2 l''} \,\la [\varphi^{(n-3)}_{0\nu_2 l} (\theta_2,\hat{\xi}_2) \otimes \varphi^{(n)}_{K'\nu_1 l} (\theta_1,\hat{\xi}_1)]_{00}
\mid 
\eol
&\times&
\left[\varphi_{0\nu'_2l''} ^{(n-3)} 
(\theta_2,\hat{\ve{\xi}}_{2})\otimes
 (1+\cos 2\theta_1)    \varphi_{K''\nu'_1l''}  ^{(n)}
(\theta_1,\hat{\xi}_{1})\right]_{00} \ra.
\eeqn{}
From (\ref{recrelphi}) one can obtain
\beq
(1+\cos 2\theta) \varphi^{(n)}_{K\nu lm} (\theta,\hat{\xi}) 
&=&
A^{(n)}_{K\nu l} \varphi^{(n)}_{K\nu+1 lm} (\theta,\hat{\xi})+
(1-B^{(n)}_{K\nu l}) \varphi^{(n)}_{K \nu lm } (\theta,\hat{\xi})
\eol
&+&
C^{(n)}_{K\nu l} \varphi^{(n)}_{K\nu-1 lm} (\theta,\hat{\xi}),
\eeqn{1+cos2t}
where
\beq
A^{(n)}_{K\nu l}= \frac{N^{(n)}_{K\nu l}}{b^{(n)}_{K\nu l}N^{(n)}_{K\nu+1 l}},
 \,\,\,\,\,\,\,\,\,\,\,
B^{(n)}_{K\nu l}= \frac{a^{(n)}_{K\nu l}}{b^{(n)}_{K\nu l}}, 
 \,\,\,\,\,\,\,\,\,\,\,
 C^{(n)}_{K\nu l} =\frac{ c^{(n)}_{K\nu l} }{ b^{(n)}_{K\nu l}} \frac{N^{(n)}_{K\nu  l}}{N^{(n)}_{K\nu-1 l}}. \,\,\,\,\,\,\,\,\,\,\,\,
 \eeqn{abc}
Then it is easy to get
\beq
(T.1) 
= \gamma^2_1 t^{(1)}_{\nu,\nu_1\nu_2 l }T^{(n)}_{\nu, \nu_1-1 \nu_2 l}, 
\eeqn{}
where
\beq
t^{(1)}_{\nu, \nu_1\nu_2,l}
=  \frac{b^{(n)}_{0\nu 0} }{b^{(n)}_{K'\nu_1-1 l}}   \frac{N^{(n)}_{0\nu+10}}{N^{(n)}_{0\nu 0}} \frac{N^{(n)}_{K'\nu_1-1  l}}{ N^{(n)}_{K'\nu_1 l}}. 
   \eeqn{Term1}

   \subsubsection{Contribution from (T.2)}

This terms comes from $4 \gamma_1 \gamma_2 \cos \theta'_1 \sin \theta'_1 \cos \theta'_2 \,(\hat{\vec{\xi}}_1 \cdot \hat{\vec{\xi}}_2)$ in $\cos 2\theta_1$.  
\beq
(T.2) 
=4\gamma_1 \gamma_2\frac{N^{(n)}_{0\nu+10}}{N^{(n)}_{0\nu 0}}b^{(n)}_{0\nu0}
\sum_{ \nu'_1 +\nu'_2+ l''=\nu} T^{(n)}_{\nu, \nu'_1 \nu'_2 l''} \, \la [\varphi^{(n-3)}_{0\nu_2 l} (\theta_2,\hat{\xi}_2) \times \varphi^{(n)}_{K'\nu_1 l} (\theta_1,\hat{\xi}_1)]_{00}
\mid 
\eol
\times
\left[\varphi_{0\nu'_2l''} ^{(n-3)} 
(\theta_2,\hat{\ve{\xi}}_{2})\times
 (\hat{\vec{\xi}}_1 \cdot \hat{\vec{\xi}}_2)\cos\theta'_1 \cos \theta'_2 \sin\theta'_1 \varphi_{K''\nu'_1l''}  ^{(n)}
(\theta_1,\hat{\xi}_{1})\right]_{00}\ra.
\eol
\eeqn{T.2}
Using 
\beq
(\hat{\vec{\xi}}_1 \cdot \hat{\vec{\xi}}_2)[Y_{l''}(\hat{\xi}_1) \times Y _{l''} (\hat{\xi}_2)]_{00}
=
- \sum_{L}\sqrt{\frac{2l''+1}{2L+1}}
(1 0 l'' 0 | L 0)^2 [ Y _{L} (\hat{\xi}_1)  \times  Y _{L}(\hat{\xi}_2) ]_{00} \eol
\eeqn{algebra}
the  (\ref{T.2}) could be further split into two distinct terms, $(T.2.-)$ and $T.2.+)$ corresponding to $L = l''-1$ and $L=l''+1$, respectively. Before proceeding to derivation each of them, it is useful to introduce the following notation:
\beq
\varphi^{(n)}_{K \nu lm}(\theta,\hat{\xi}) = \phi^{(n)}_{K \nu l}(\theta) Y_{lm}(\hat{\xi})
\eeqn{varphi}
\\
\\
{\it Contribution from $(T.2.-)$}
\vspace{0.3 cm}
\\
When developing this term  the following matrix element was used:
\beq
& \la & \varphi^{(n-3)}_{0 \nu_2 lm}(\theta_2,\hat{\xi}_2) \mid \cos \theta_2 \, \phi^{(n-3)}_{0 \nu'_2 l''} (\theta_2) \, Y_{l''-1 m}(\hat{\xi}_2) \ra
\eol
&=&
\delta_{l''-1, l} \delta_{\nu'_2 , \nu_2} \delta_{K'',K'+1}
\frac{ N^{(n-3)}_{0\nu_2 l+1}}{N^{(n-3)}_{0\nu_2 l }}
 \frac{\nu_2+l+3/2}{2\nu_2+(n-5)/2+l+1}
\eol
&+&
\delta_{l''-1, l} \delta_{\nu'_2+1, \nu_2}\delta_{K'',K'-1}\frac{ N^{(n-3)}_{0\nu_2-1 l+1}}{N^{(n-3)}_{0\nu_2 l }} \frac{\nu_2}{2\nu_2+(n-5)/2+l -1}, 
\eeqn{ME_theta_2}
which could be easily obtained from recurrence relations between Jacobi polynomials with different nodes numbers but the same other parameters. After this one has  to integrate the product of  $\cos \theta_1  \sin \theta_1$ and the hyperradial functions   $(\theta_1,\hat{\xi}_1)$. This requires special attention because using the same recurrence relations leads to the matrix elements of the type $\la \phi^{(n)}_{K \nu l} (\theta_1,\hat{\xi}_1) \mid \phi^{(n)}_{K' \nu l}(\theta_1,\hat{\xi}_1)\ra$ which are  not equal to $   \delta_{KK'}$. Therefore, the $\phi^{(n)}_{K'' \nu'_1 l''}  (\theta_1)$ was rearranged into the sum of the hyperangular functions  $\phi^{(n)}_{K'  \nu''_1 l }  (\theta_1)$ over $\nu''_1$. 
In the $K''=K'+1$ case it results in 
\beq
\la \varphi^{(n)}_{K'\nu_1 l} (\theta_1,\hat{\xi}_1)\mid \cos \theta_1  \sin \theta_1 \, \phi^{(n)}_{K'+1 \nu'_1 l+1}  (\theta_1)\, Y_{l  m }(\hat{\xi}_1)\ra 
\eol
=
- \delta_{\nu'_1+2,\nu_1}
\frac{N^{(n)}_{K'+1 \,\nu_1-2 \,l+1}}{N^{(n)}_{K'  \nu_1 l }}
\frac{4(\nu_1-1)\nu_1}{(  n +4\nu  )({ n+4\nu-2})} + ...,
\eeqn{ME_theta_1}
where by $...$ the terms are denoted that do not satisfy $\nu_1+\nu_2+l=\nu+1$ and, therefore, will not contribute to $(T.2.-)$. For the $K''=K'-1$ case, the following was obtained,
\beq
\la \varphi^{(n)}_{K'\nu_1 l} (\theta_1,\hat{\xi}_1)|\cos \theta_1  \sin \theta_1 \phi^{(n)}_{K'-1 \nu'_1 l+1}  (\theta_1)Y_{l  m }(\hat{\xi}_2)\ra 
\eol
=
 \delta_{\nu'_1+1,\nu_1}
\frac{N^{(n)}_{K'-1 \nu_1-1 l+1}}{N^{(n)}_{K'  \nu_1  l }}
\frac{\nu_1(2\nu +  n/2 -\nu_1)}{({2\nu + n/2-1 })(2\nu+n/2)} + ...
\eeqn{ME_theta_1}
where once again by $...$ the terms are denoted that do not satisfy $\nu_1+\nu_2+l=\nu+1$ and that don't contribute to $(T.2.-)$.
Collecting everything together, one obtains 
\beq
(T.2-) 
=
- \gamma_1 \gamma_2 \left(t^{(2-+)}_{\nu,\nu_1 \nu_2 l} T^{(n)}_{\nu \nu_1-2 \nu_2 l+1}+ t^{(2--)}_{\nu,\nu_1 \nu_2 l} T^{(n)}_{\nu \nu_1-1 \nu_2-1 l+1}\right)
\eeqn{T2-}
where
\beq
t^{(2-+)}_{\nu,\nu_1 \nu_2 l}
&=&
-\frac{l+1}{\sqrt{ (2l+3)(2l+1)}}
  \frac{\nu_2+l+3/2}{2\nu_2+(n-5)/2+l+1} 
\frac{4(\nu_1-1)\nu_1}{(  n +4\nu  )({ n+4\nu-2})}
\eol
&\times&
\frac{N^{(n)}_{0\nu+10}}{N^{(n)}_{0\nu 0}}b^{(n)}_{0\nu0} 
\frac{ N^{(n-3)}_{0\nu_2 l+1}}{N^{(n-3)}_{0\nu_2 l }}
\frac{N^{(n)}_{K'+1 \,\nu_1-2 \,l+1}}{N^{(n)}_{K'  \nu_1 l }} 
\eeqn{t2-+}
and
\beq
t^{(2--)}_{\nu,\nu_1 \nu_2 l} 
&=&
 \frac{l+1}{\sqrt{ (2l+3)(2l+1)}}
\frac{4\nu_1(2\nu +  n/2 -\nu_1)}{({2\nu + n/2-1 })(2\nu+n/2)} \frac{\nu_2}{2\nu_2+(n-5)/2+l -1} 
\eol
&\times&
\frac{N^{(n)}_{0\nu+10}}{N^{(n)}_{0\nu 0}}b^{(n)}_{0\nu0} 
  \frac{ N^{(n-3)}_{0\nu_2-1 l+1}}{N^{(n-3)}_{0\nu_2 l }} 
\frac{N^{(n)}_{K'-1 \nu_1-1 l+1}}{N^{(n)}_{K'  \nu_1  l }}.
\eeqn{t2--}
\\
\\
{\it Contribution from $(T.2.+)$}
\vspace{0.3 cm}
\\
To develop this term the following is used:
\beq
& \la& \varphi^{(n-3)}_{0 \nu_2 lm}(\theta_2,\hat{\xi}_2) \mid \cos \theta_2\,  \phi^{(n-3)}_{0 \nu'_2 l''} (\theta_2)\, Y_{l''+1 m}(\hat{\xi}_2) \ra
\eol
&=&
\delta_{l''+1, l} \delta_{\nu'_2 , \nu_2} \delta_{K'',K'-1}
\frac{ N^{(n-3)}_{0\nu_2 l-1}}{N^{(n-3)}_{0\nu_2 l }}
 \frac{\nu_2+(n-5)/2+l-1}{2\nu_2+(n-5)/2+l-1}
\eol
&+&
\delta_{l''+1, l} \delta_{\nu'_2-1, \nu_2}\delta_{K'',K'+1}\frac{ N^{(n-3)}_{0\nu_2+1 l-1}}{N^{(n-3)}_{0\nu_2 l }} \frac{\nu_2+(n-5)/2-1/2}{2\nu_2+(n-5)/2+l +1} 
\eeqn{ME_theta_2}
As in the previous case of $(T.2.-)$ a special treatment is needed for the matrix element involving hyperangular functions of $(\theta_1,\hat{\xi}_1)$. 
A strategy similar to the one above, gives for  $K''=K'-1$  
\beq
\la \varphi^{(n)}_{K'\nu_1 l} (\theta_1,\hat{\xi}_1)|\cos \theta_1  \sin \theta_1 \phi^{(n)}_{K'-1 \nu'_1 l-1}  (\theta_1)Y_{l  m }(\hat{\xi}_2)\ra 
\eol
=
\delta_{\nu'_1,\nu_1}
\frac{N^{(n)}_{K'-1 \nu'_1 l-1}}{N^{(n)}_{K'  \nu'_1  l }}
\left( 1 -\frac{2\nu_1}{n+4\nu-2}\right)\left(1-\frac{2\nu_1}{n+4\nu }\right) + ...
\eeqn{ME_T2+-}
while for $K''=K'+1$ 
\beq
\la \varphi^{(n)}_{K'\nu_1 l} (\theta_1,\hat{\xi}_1)\mid \cos \theta_1  \sin \theta_1 \,\phi^{(n)}_{K'+1 \nu'_1 l-1}  (\theta_1)\,Y_{l  m }(\hat{\xi}_1)\ra 
\eol
=
-\delta_{\nu'_1+1,\nu_1}\frac{N^{(n)}_{K'+1 \nu_1-1 l-1}}{N^{(n)}_{K'  \nu_1  l }}\frac{2\nu_1 }{n+4\nu  -2}
\frac{n+4\nu-2\nu_1 }{n+4\nu }+ ...
\eeqn{ME_T2++}
Both in (\ref{ME_T2+-}) and (\ref{ME_T2++}) by ... the terms are denoted that do not satisfy $\nu_1+\nu_2+l = \nu+1$ and, therefore, do not contribute to $(T.2.+)$.
Then, collecting everything together, one obtains
\beq
(T.2.+)  
 =
 \gamma_1 \gamma_2 \left(t^{(2+-)}_{\nu,\nu_1 \nu_2 l}T^{(n)}_{\nu,  \nu_1 \nu_2 l-1}+t^{(2++)}_{\nu,\nu_1 \nu_2 l}T^{(n)}_{\nu,  \nu_1-1 \nu_2+1 l-1} \right),
\eeqn{T2+}
where
\beq
t^{(2+-)}_{\nu,\nu_1 \nu_2 l} &=& 
- 4 b^{(n)}_{0\nu0} \frac{l}{\sqrt{(2l-1)(2l+1)}} \frac{N^{(n)}_{0\nu+10}}{N^{(n)}_{0\nu 0}} 
\frac{ N^{(n-3)}_{0\nu_2 l-1}}{N^{(n-3)}_{0\nu_2 l }}\frac{N^{(n)}_{K'-1 \nu_1 l-1}}{N^{(n)}_{K'  \nu_1  l }}
\eol
&\times&
\left(1-\frac{2\nu_2}  {n+4\nu_2 +2l-7}\right)
\left( 1 -\frac{2\nu_1}{n+4\nu-2}\right)\left(1-\frac{2\nu_1}{n+4\nu }\right) \,\,\,\,\,\,\,\,\,\,\,\,\,
\\
t^{(2++)}_{\nu,\nu_1 \nu_2 l} &=& 
 4 b^{(n)}_{0\nu0} 
\frac{l}{\sqrt{(2l-1)(2l+1)}} \frac{N^{(n)}_{0\nu+10}}{N^{(n)}_{0\nu 0}}
\frac{ N^{(n-3)}_{0\nu_2+1 l-1}}{N^{(n-3)}_{0\nu_2 l }} \frac{N^{(n)}_{K'+1 \nu_1-1 l-1}}{N^{(n)}_{K'  \nu_1  l }}
\eol
&\times&
\left(1- \frac{\nu_2 +l+3/2}{2\nu_2+(n-3)/2+l } \right)\frac{2\nu_1 }{n+4\nu  -2}
\frac{n+4\nu-2\nu_1 }{n+4\nu }. 
\eeqn{t2+-}

\subsubsection{ Contribution from $(T.3)$}

This terms arises from the $2\gamma^2_2\cos^2\theta'_2 \sin^2 \theta'_2$ part of $\cos 2\theta_1$ and it is given by the expression
\beq
(T.3)
=
\gamma^2_2 b^{(n)}_{0\nu 0} \frac{N^{(n)}_{0\nu+10}}{N^{(n)}_{0\nu 0}} \sum_{ \nu'_1+ \nu'_2 +l''=\nu}
T^{(n)}_{\nu,\nu'_1 \nu'_2 l''}
\la [\varphi^{(n-3)}_{0\nu_2 l} (\theta'_2,\hat{\xi}'_2) \times \varphi^{(n)}_{K'\nu_1 l} (\theta'_1,\hat{\xi}'_1)]_{00}
\mid 
 \, 
\eol
\times
\left[ (1+\cos 2\theta'_2)\,\varphi_{0\nu'_2l''} ^{(n-3)} (\theta'_2,\hat{\ve{\xi}}'_{2})\times
 \sin^2\theta'_1
 \varphi_{K''\nu'_1l''}  ^{(n)}
(\theta'_1,\hat{\xi}'_{1})\right]_{00}. \ra
\eol
\eeqn{T3}
After applying (\ref{1+cos2t}) to $(1+\cos 2\theta'_2)\,\varphi_{0\nu'_2l''} ^{(n-3)} (\theta'_2,\hat{\ve{\xi}}'_{2})$ the $(T.3)$ splits further into  three terms that contain $ \la \phi^{(n)}_{K'\nu_1 l} (\theta'_1)\mid  \sin^2\theta'_1 |
 \phi_{K''\nu'_1l}  ^{(n)}(\theta'_1) \ra$ with $K''=K'-2, K'$ and $K'+2$. To deal with them the following relation is used:
\beq
(1-\cos 2\theta) \varphi^{(n)}_{K\nu lm} (\theta,\hat{\xi}) 
&=&
-A^{(n)}_{K\nu l} \varphi^{(n)}_{K\nu+1 lm} (\theta,\hat{\xi})+
\left[1+B^{(n)}_{K\nu l}\right] \varphi^{(n)}_{K \nu lm } (\theta,\hat{\xi}) 
\eol
&-&
C^{(n)}_{K\nu l} \varphi^{(n)}_{K\nu-1 lm} (\theta,\hat{\xi}).
\eeqn{one-sinus}
Applying this relation for $K''=K$ gives the immediate result  for   its contribution to $(T.3)$, which contains  the same  quantity $t^{(1)}_{\nu,\nu_1 \nu_2 l}$ that is present in $(T.1)$. In the case of $K''=K'-2$   relation (\ref{one-sinus}) is applied to $ \la \phi^{(n)}_{K'\nu_1 l} (\theta'_1)   \sin^2\theta'_1 \mid$ and then $K'-2$ in $\mid \phi_{K'-2 \nu'_1l}  ^{(n)}(\theta'_1) \ra$ is raised by two units with the help of recurrence relations for Jacobi polynomials. This allows the orthogonality relation (\ref{ortho}) to be used. In a similar way,  in the case of $K''=K'+2$ relation (\ref{one-sinus}) is applied to $\mid \sin^2\theta'_1 \phi_{K'+2 \nu'_1l}  ^{(n)}(\theta'_1) \ra$ and then the $K'$ in $ \la \phi^{(n)}_{K'\nu_1 l} (\theta'_1) \mid$ is raised by two units.
The final result of these actions is
\beq
(T.3)
&=& \gamma^2_2 t^{(3.1)}_{\nu,\nu_1 \nu_2 l} T^{(n)}_{\nu,\nu_1 \nu_2-1 l}-\frac{1}{2} \gamma^2_2 \left(1 - B^{(n-3)}_{0 \nu_2 l}\right) t^{(1)}_{\nu, \nu_1 \nu_2 l}T^{(n)}_{\nu, \nu_1-1 \nu_2 l}
\eol
&+&\gamma^2_2  t^{(3.3)}_{\nu,\nu_1 \nu_2 l} T^{(n)}_{\nu, \nu_1-2 \, \nu_2+1 \, l},
\eeqn{T3final}
where
\beq
t^{(3.1)}_{\nu,\nu_1 \nu_2 l} 
&=&
\frac{(\nu_1 + K'+n/2+l-3)_2}{(2\nu_1 + K'+n/2+l-3)_2} 
b^{(n)}_{0\nu 0} A^{(n-3)}_{0 \nu_2-1 l}\frac{N^{(n)}_{0\nu+10}}{N^{(n)}_{0\nu 0}} 
\,
\frac{N^{(n)}_{K'-2 \nu_1 l}}{N^{(n)}_{K'  \nu_1 l}} \,\,\,\,\,\,\,\,\,\,\,\,\,\,\,\,
\eeqn{t2}
and
\beq
t^{(3.3)}_{\nu,\nu_1 \nu_2 l} & =&    b^{(n)}_{0 \nu 0}\frac{N^{(n)}_{0 \nu+1 0}}{N^{(n)}_{0 \nu 0}} \,\frac{N^{(n)}_{2\nu_2+l+2, \nu_1-2 l}}{N^{(n)}_{2\nu_2+l, \nu_1 l}} \, \frac{\nu_1(\nu_1-1)}{(2\nu+ n/2)(2\nu+n/2-1)}C^{(n-3)}_{0 \nu_2+1 \, l}. \eol
\eeqn{T33}


\subsection{Matrix elements for hyperradial potentials}
\label{sec:C}
\setcounter{equation}{0}


\subsubsection{Term $V^{(1)}_{\rm dir}$}

This term contains interactions of bosons within the $A-1$ core. There are two ways of dealing with it. 
\\
(1)
An integration over the hyperspherical angles of the $A-1$ system could be done first to get $\la Y_0(1,2,...,A-1) \mid \sum_{i<j}^{A-1} V_{ij} \mid Y_0(1,2,...,A-1) \ra  = V_{A-1}(\rho \sin \theta)$, which for the gaussian interaction (\ref{2bpot}) is described by the confluent hypergeometric function    $ M(\frac{3}{2},\frac{n-3}{2}, -\frac{2\rho^2 \sin^2 \theta}{a^2})$ \cite{Tim12}. Then second integration should be performed over the  angle $\theta$. This will not contain any $\delta_{\nu \nu'}$. 
\\
(2)
An alternative approach, used for all numerical calculations here,  involves a different choice of the hyperspherical coordinate $\theta_2 = \frac{1}{\sqrt{2}}|\ve{r}_{A-1} - \ve{r}_{i} |/ \rho_c$, where $\rho_c = \rho \sin \theta_1$ while keeping the same choice of $\cos \theta_1$. In these coordinates the two-body potential $V_{A-1,i}$ between  $A-1$ and $i$, both belonging to the $A-1$ core, is a function of $\sqrt{2} \rho \sin \theta_{1} \cos \theta_{2}$.
After the  
 transformation $(\gamma_1=0, \gamma_2=1)$ to new hyperspherical coordinates defined by $\vec{\xi}'_{1} = \vec{\xi}_{2}$ and $\vec{\xi}'_{2} =- \vec{\xi}_{1}$ the  $V_{A-1,i}$ depends on $\cos \theta'_{1}$ only, which  reduces the task of calculating $V_{\rm dir}^{(1)}$ to performing a single integration.
This results in the following expression $V^{(1)}_{\rm dir}$:
\beq
V^{(1)}_{\rm dir; \nu \nu'}(\rho) 
&=& \frac{(A-1)(A-2)}{2}\left( {\cal N}_{\nu}{\cal N}_{\nu'}\right)^{-1}
\sum_{\nu_2=0}^{\min(\nu,\nu')} \delta_{\nu_1,\nu-\nu_2}\delta_{\nu'_1,\nu'-\nu_2}
 \eol
& \times &
 T^{(n)}_{\nu,\nu_1 \nu_2 0}(0,1)
 \,T^{(n)}_{\nu',\nu'_1 \nu_2 0}(0,1)
\, v^{(n)}_{2\nu_2 \nu'_1  0;\,2\nu_2  \nu_1  0}(\rho),
\eeqn{Vdir1}
where
\beq
 v^{(n)}_{K' \nu'   l ;K' \nu   l}(\rho)= \la \phi^{(n) }_{K' \nu'  l }(\cos \theta)
| V(\rho \cos \theta) |
 \phi^{(n) }_{K' \nu l }(\cos \theta)\ra.
 \eeqn{v}

\subsubsection{Terms $V^{(2)}_{\rm dir}$ and $V^{(1)}_{\rm ex}$}
The second part $V^{(2)}_{\rm dir}$ of the direct  and the first part $V^{(1)}_{ex}$ of the exchange contributions to the hyperradial potential, given by (\ref{Vdir}) and (\ref{ex}) are very similar. They contain the interaction of the bosons with numbers $A-1$ and $A$. Extracting boson $A-1$ from $Y_0$ in $V^{(2)}_{\rm dir}$,   boson  $A$ from $Y_0(1,...,A-2,A)$ and bosons $A-1$ from $Y_0(1,...,A-2,A-1)$ in $V^{(1)}_{\rm ex}$,   followed by the coordinate transformation shown in Fig. 1 one obtains
\beq
V_{{\rm dir};\nu \nu'}^{(2)}+(A-1)V_{{\rm ex}; \nu \nu'}^{(1)} &=&
(A-1) \left( {\cal N}_{\nu}{\cal N}_{\nu'}\right)^{-1}
\eol
&\times &
\sum_{\nu_2    l}\left[1+(-)^{l}\right]\delta_{\nu_1, \nu-\nu_2-l}\delta_{\nu'_1, \nu-\nu'+\nu'_1} \delta_{K',2\nu_2+l}\,\, 
 \eol
& \times&
T^{(n)}_{\nu;  \nu_1 \nu_2 l}(\gamma^{(1)}_1, \gamma^{(1)}_2)\,\, T^{(n)}_{ \nu'; \nu'_1 \nu_2 l} (\gamma^{(1)}_1, \gamma^{(1)}_2) \,\,
 v^{(n)}_{K' \nu'_1   l ;K' \nu_1   l}(\rho)
\eol
 \eeqn{}
with
\beq
 \gamma^{(1)}_1 = \sqrt{\frac{1}{2}\frac{A }{(A-1) }} , 
 \,\,\,\,\,\,\,\,\,\,\,\,
 \gamma^{(1)}_2 =   \sqrt{\frac{1}{2} \frac{A-2 }{A-1}}
 \eeqn{gam_(1)}
 
 \subsubsection{Terms $V^{(2)}_{\rm ex}$ and $V^{(3)}_{\rm ex}$.}
 
\begin{figure}[t]
\vspace{0.5 cm}
\includegraphics[scale=0.38]{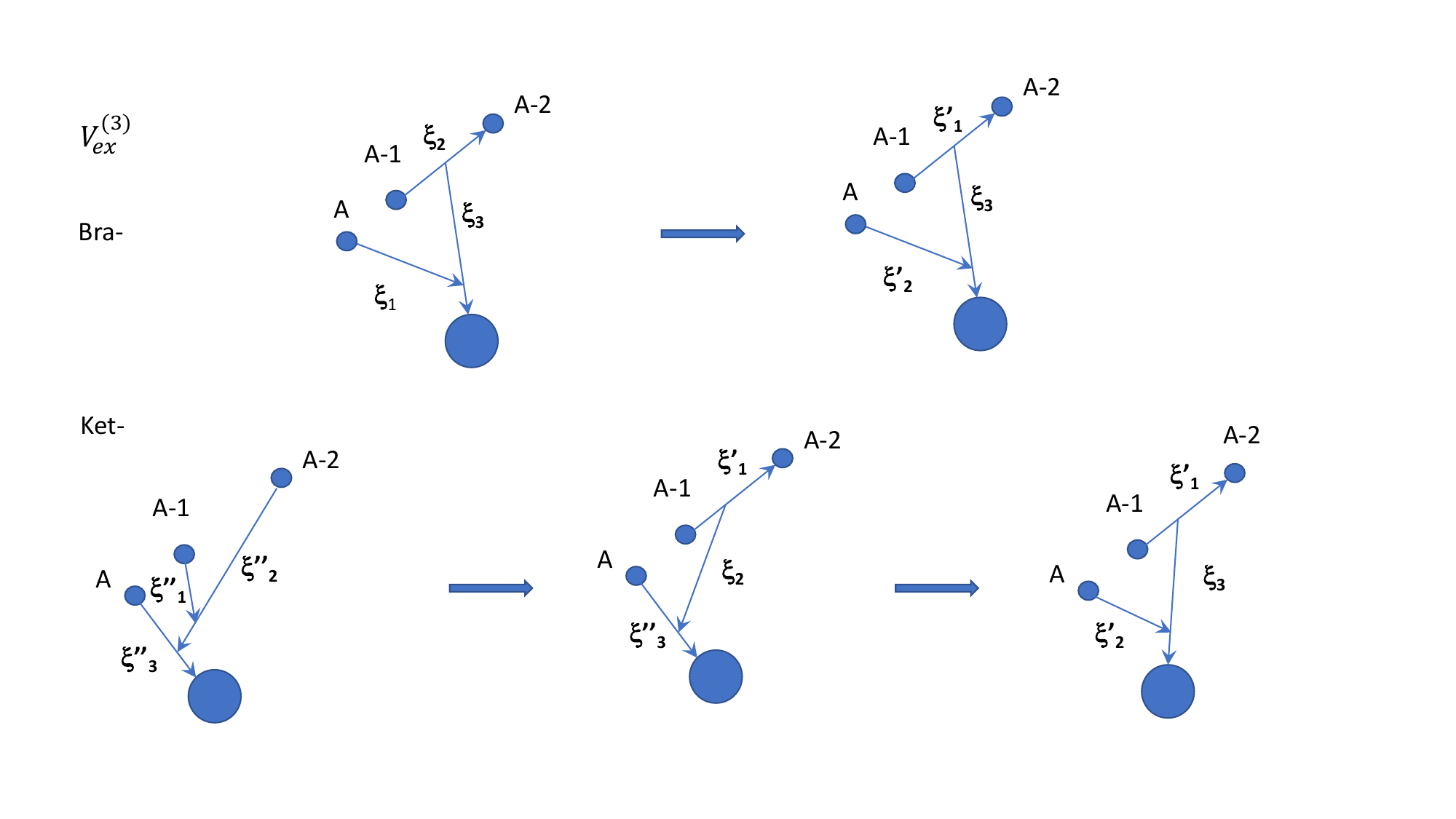}
\caption{The coordinate transformations for calculating $V^{(3)}_{\rm dir}$ }
\label{fig:V23ex}
\end{figure}

 The exchange terms $V_{ex}^{(2)}$ and $V_{ex}^{(3)}$ are similar. Their sum can be written as
\beq
& &V_{{\rm ex}; \nu \nu'}^{(2)} +V_{{\rm ex}; \nu \nu'}^{(3)} = 
(A-2) \left( {\cal N}_{\nu}{\cal N}_{\nu'}\right)^{-1}
\eol
&\times&
{\big [}\la {  Y}_{0} (1,...,A-2,A) \varphi^{(n)}_{0\nu0} (A-1)|V_{A-2,A}|
{  Y}_{0} (1,...,A-2,A-1)\varphi^{(n)}_{0 \nu' 0}(A )\ra 
\eol
 &+ &\la {  Y}_{0} (1,...,A,A-2)  \varphi^{(n)}_{0\nu0} (A-1)|V_{A-2,A-1}|
{  Y}_{0} (1,...,A-1,A-2)\varphi^{(n)}_{0 \nu' 0}(A )\ra \big].
\eol
\eeqn{}
Swapping $A$ and $A-1$ in the notation of the second term one can see that $V_{{\rm ex}; \nu \nu'}^{(2)}= V_{{\rm ex}; \nu'\nu}^{(3)}$. The calculation of any of them  is preformed by separating the hyperangular functions of bosons $A$, $A-1$ and $A-2$ in the HCHs followed by coordinate transformations, shown in Fig. \ref{fig:V23ex}.
  Then
 \beq
&& V_{ex; \nu'\nu}^{(2)} (\rho) +V_{{\rm ex}; \nu \nu'}^{(3)}= (A-2)
 \sum_{\nu_2=0}^{\nu} \, T^{(n-3)}_{\nu_2,\nu_2 \, 0 0}(\gamma^{(2)}_1,\gamma^{(2)}_2)
\eol
&\times &
{\big [}
\ T^{(n)}_{\nu',\nu'_1 \nu_2 0}(0,1) \, 
 T^{(n)}_{\nu,\nu_1 \nu_20 }(\gamma^{(1)}_1,\gamma^{(1)}_2)
 +
 T^{(n)}_{\nu',\nu'_1 \nu_2 0}(0,1) \, 
 T^{(n)}_{\nu,\nu_1 \nu_20 }(\gamma^{(1)}_1,\gamma^{(1)}_2) {\big]}
 \eol
 &\times&
 \delta_{\nu_1, \nu-\nu_2} \delta_{\nu'_1, \nu'-\nu_2}
 v^{(n)}_{K' \nu'_1   l ;K' \nu_1   l}(\rho),
 \eeqn{Vex23}
where the $\gamma^{(2)}_{1,2}$ are:
\beq
\gamma^{(2)}_1 = -\sqrt{\frac{2}{(A-2)(A-1)}}, \,\,\,\,\,\,\,\,\,\,\,
\gamma^{(2)}_2 =  \sqrt{\frac{(A-3)A}{(A-2)(A-1)}}
\eeqn{gammas2}

 \subsubsection{Term $V^{(4)}_{\rm ex}$ }

The calculation of the final exchange term in the hyperradial potential,
\beq
& &V_{{\rm ex}; \nu \nu'}^{(4)}  = 
\frac{(A-2)(A-3)}{2} \left( {\cal N}_{\nu}{\cal N}_{\nu'}\right)^{-1}
\eol
&\times&
\la {  Y}_{0} (1,...,A-2,A) \varphi^{(n)}_{0\nu0} (A-1)|V_{A-2,A-3}|
{  Y}_{0} (1,...,A-2,A-1)\varphi^{(n)}_{0 \nu' 0}(A )\ra ,
\eol
 \eeqn{Vex4}
is based on separating the hyperangular functions of bosons $A$, $A-1$, $A-2$ and $A-3$ and making the coordinate transformations shown in Fig. \ref{fig:V4ex}. This results in
\beq
V^{(4)}_{{\rm ex}; \nu'\nu}& =& \frac{(A-2)(A-3)}{2}
\sum_{\nu_2=0}^{\min(\nu,\nu')} T^{(n)}_{\nu,\nu_1 \nu_2 0}(0,1)
 T^{(n)}_{\nu',\nu'_1 \nu_2 0}(0,1) 
 \, T^{(n-3)}_{\nu_2,\nu_2 0 0}(\gamma^{{\cal N}}_1, \gamma^{{\cal N}}_2) 
\eol
&\times&
\delta_{\nu_1, \nu-\nu_2}  \delta_{\nu'_1, \nu'-\nu_2}
 v^{(n)}_{K' \nu'_1   l ;K' \nu_1   l}(\rho).
\eeqn{V4ex}

\begin{figure}[t]
\vspace{0.4 cm}
\includegraphics[scale=0.36]{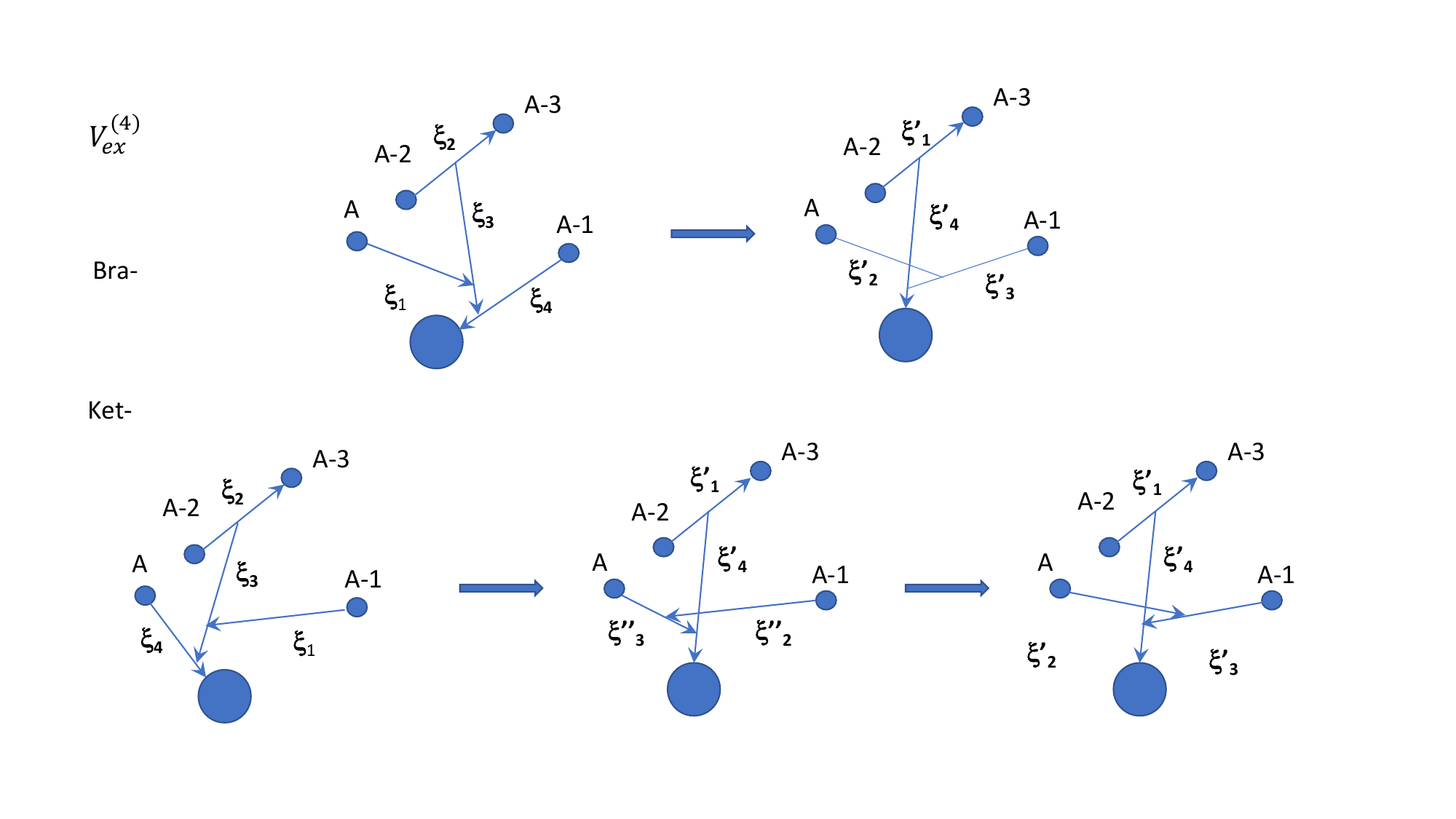}
\caption{The coordinate transformations for developing $V^{(4)}_{{\rm ex}}$}
\label{fig:V4ex}
\end{figure}

\begin{acknowledgements}
This work was supported by the United Kingdom Science and Technology Facilities Council (STFC) under Grant  No.   ST/V001108/1.
\end{acknowledgements}
\nocite{*}


\end{document}